\documentclass{iopart}
\usepackage{iopams}
\usepackage{graphicx}
\usepackage{amsthm}
\usepackage{mmat} 
\usepackage{mxarrow} 

\newtheorem{alg}{Algorithm}
\newtheorem{defn}{Definition}
\def\theequation{{\arabic{section}}.{\arabic{equation}}}

\begin{document}
\title{Point source identification in non-linear advection-diffusion-reaction systems}
\author{A.V. Mamonov$^1$ and Y.-H. R. Tsai$^2$}
\address{$^1$Institute for Computational Engineering and Sciences (ICES),
The University of Texas at Austin, 201 East 24th St Stop C0200, Austin, TX 78712 USA}, 
\address{$^2$Department of Mathematics and ICES, 
The University of Texas at Austin,\\ 1 University Station C1200, Austin, TX 78712 USA}
\eads{\mailto{mamonov@ices.utexas.edu} and \mailto{ytsai@math.utexas.edu}}

\begin{abstract}
We consider a problem of identification of point sources in time dependent advection-diffusion systems with a 
non-linear reaction term. The linear counterpart of the problem in question can be reduced to solving a system of 
non-linear algebraic equations via the use of adjoint equations. We extend this approach by constructing an algorithm 
that solves the problem iteratively to account for the non-linearity of the reaction term. We study the question 
of improving the quality of source identification by adding more measurements adaptively using the solution obtained 
previously with a smaller number of measurements. 
\end{abstract}

\section{Introduction}
\label{sec:intro}

We are interested in a problem of identification of point sources in non-linear time dependent 
advection-reaction-diffusion systems from a sparse set of measurements. Here by sparse we mean a small number of spatially separated measurements. This work is motivated by applications in atmospheric studies where one
would like to localize a release of an airborne contaminant \cite{atmoinv, enting2002inverse}. 
A possible model for such problem is a linear scalar parabolic equation with a known first order advection 
term and point sources \cite{akcelik2006inversion}. However, for more realistic modeling of the processes in 
the atmosphere one needs to consider a system with multiple chemical species that react with each other. 
In some cases it may even be beneficial to make measurements not of the concentration of the contaminant itself, 
but of the products of its reactions with the other species in the atmosphere. This leads to studying not just 
a single parabolic equation, but a system of such equations. Moreover, an accurate modeling of the chemical 
reactions between the different chemical species requires the use of non-linear reaction terms 
\cite{jacobson2005fundamentals, kim1997computation, sportisse2000analysis} of large magnitudes that lead to 
very stiff systems. To our knowledge this is the first study of the source identification problem for non-linear 
advection-diffusion-reaction systems.

To solve the source identification problem with sparse measurements one needs to assume some sparsity of the 
unknown source term as well. Here we assume that both sources and measurements are point-like. Under sparsity
assumptions in the linear case the source identification problem may be reduced to solving a system of algebraic
equations obtained by employing the relation between the forward model and its adjoint 
\cite{burger2009discovering, fhe2011, kasibhatla2000inverse}. The adjoint problem solution is not coupled to the (unknown) forward problem solution, which makes the problem much easier to solve numerically compared to general 
PDE constrained optimization problems that arise if no sparsity constraints are used. What complicates the 
non-linear case is that the forward and adjoint solutions are no longer uncoupled. In this work we propose 
a computationally efficient iterative procedure that resolves this coupling and solves the source identification 
problem simultaneously. 
Note that one can try to exploit the sparse nature of sources and measurements to use the ideas from compressed 
sensing \cite{candes2006robust, candes2006stable, yin2008bregman} to recover the sources \cite{li2011heat}. 
This approach can be beneficial if the number of point sources in the system is large. However, the main idea 
of compressed sensing of replacing $L_0$ optimization with $L_1$ optimization requires some properties of the 
forward operator (like the restricted isometry property), which the forward parabolic operator may not satisfy. 
Thus, we use a different approach here.

Another aspect of source identification that we consider here is an efficient placement of measurements. In the 
presence of noise in the measured data or some uncertainty in the system's parameters an efficient placement of 
the measurements in the domain of interest may play a crucial role in stable source identification. Here we consider both a priori placement of initial measurements, when one has no prior knowledge about the possible source 
distribution, and a posteriori placement of additional measurements, when one utilizes the source estimate obtained 
with a fewer measurements to place new ones. The problem of efficient positioning of measurements is known in the
literature under the name experimental design or optimal design of experiments 
\cite{pukelsheim2006optimal, haber2008numerical}. Here we propose a heuristic for adaptive placement of new 
measurements based on the study of a single source case. It is not optimal in the sense that it relies on making
redundant measurements, however it is computationally inexpensive and it performs well in the numerical experiments 
that we consider.

\section{Non-linear advection-reaction-diffusion system with point sources}
\label{sec:system}

A general parabolic system of equations with $n$ components 
$$\boldsymbol u(\boldsymbol x, t) = (u_1(\boldsymbol x, t), \ldots, u_n(\boldsymbol x, t))^T$$
studied here has the form
\begin{equation}
\boldsymbol u_t = 
\boldsymbol D \boldsymbol \Delta \boldsymbol u  
- \boldsymbol w \cdot \boldsymbol \nabla \boldsymbol u
+ \boldsymbol L \boldsymbol u
+ \boldsymbol Q(\boldsymbol u) \boldsymbol u
+ \boldsymbol f, \quad \boldsymbol x \in \Omega, \quad t \in [0, T],
\label{eqn:usystem}
\end{equation}
for some domain $\Omega \subset \mathbb{R}^d$ and terminal time $T > 0$. The methods presented here are applicable 
for any $d \geq 1$, and while the most relevant case for applications is $d=3$, for the ease of visualization 
we consider examples in $d=1,2$ dimensions. Hereafter bold lowercase letters denote vectors and vector-functions 
and bold uppercase letters denote matrices and matrix-functions.  Dirichlet and Neumann conditions are specified 
on the corresponding parts of the boundary
\begin{equation}
\left. \boldsymbol u \right|_{\Gamma_D} = \boldsymbol u_D, \quad
\left. \frac{\partial \boldsymbol u}{\partial \nu} \right|_{\Gamma_N} = \boldsymbol \psi, \quad
\partial \Omega = \Gamma_D \cup \Gamma_N,
\label{eqn:ubc}
\end{equation}
and the initial condition is 
\begin{equation}
\boldsymbol u(\boldsymbol x, 0) = 0. 
\label{eqn:uic}
\end{equation} 
The diffusion and advection terms are given in terms of the diagonal matrices
\begin{equation}
\fl
\boldsymbol D = \begin{bmatrix}
\epsilon_1 &        & 0 \\
    & \ddots & \\
0   &        & \epsilon_n
\end{bmatrix}, \quad
\boldsymbol \Delta = \begin{bmatrix}
\Delta &        & 0 \\
           & \ddots & \\
0          &        & \Delta
\end{bmatrix}, \quad
\boldsymbol \nabla = \begin{bmatrix}
\nabla &        & 0 \\
           & \ddots & \\
0          &        & \nabla
\end{bmatrix},
\end{equation}
where $\epsilon_j > 0$, $j=1,\ldots,n$ are \emph{diffusion constants} and 
$\boldsymbol w(\boldsymbol x): \Omega \to \mathbb{R}^d$ is the vector \emph{advection field}. The dot product
$\boldsymbol w \cdot \boldsymbol \nabla$ in (\ref{eqn:usystem}) is understood componentwise, i.e.
\begin{equation}
\boldsymbol w \cdot \boldsymbol \nabla \boldsymbol u = 
\mbox{diag} \left( \boldsymbol w \cdot \nabla u_1, \ldots, \boldsymbol w \cdot \nabla u_n \right).
\end{equation}
Note that the diffusion and advection terms are linear operators. The only source of non-linearity in the system is 
the \emph{reaction term}
\begin{equation}
\boldsymbol R(\boldsymbol u) = \boldsymbol L \boldsymbol u + \boldsymbol Q(\boldsymbol u) \boldsymbol u,
\end{equation}
which we split into the linear $\boldsymbol L$ and non-linear $\boldsymbol Q(\boldsymbol u)$ parts. 

We consider the source terms of the form
\begin{equation}
f_k (\boldsymbol x, t) = \sum_{j = l_k+1}^{l_{k+1}} a_j h_j(t) \delta(\boldsymbol x - \boldsymbol y^j), 
\quad k = 1, \ldots, n,
\label{eqn:fsrc}
\end{equation}
where the time dependent part $h_j(t)$ of the source term is either a point source $\delta (t - \tau_j)$ or 
an indicator function of some time interval. In the simplest case it is an indicator function of $[0,T]$. 
The source \emph{intensities} $a_j \geq 0$ are assumed to be constant in time. The spatial location of $j^{th}$ 
source is $\boldsymbol y^j \in \Omega$. The parameters $0 = l_1 \leq l_2 \leq \ldots \leq l_n \leq l_{n+1} = N_s$ 
determine the number of sources in each component, which is $l_{k+1} - l_k$. The total number of sources in the
system is denoted by $N_s$.

Existence and uniqueness of solutions of non-linear elliptic and parabolic systems is typically established using a 
fixed point iteration technique \cite{smoller1994shock}. For example, in case of a scalar elliptic equation
\begin{equation}
A u + R(u) + f(x) = 0, \quad \boldsymbol x \in \Omega,
\label{eqn:elliptic}
\end{equation}
with an elliptic operator $A$ and a non-linear reaction term satisfying
\begin{equation}
\frac{\partial R}{\partial u} + \kappa > 0, \quad (\boldsymbol x,u) \in \overline{\Omega} \times [m,M], \quad
\kappa,m,M > 0,
\end{equation}
the iteration
\begin{equation}
(A - \kappa) u^{q+1} = - \left( R(u^q) + f(\boldsymbol x) + \kappa u^q \right), \quad q=0,1,2,\ldots
\end{equation}
has a unique fixed point that is a solution of (\ref{eqn:elliptic}) \cite{smoller1994shock}. Similar results can be 
obtained for parabolic non-linear systems. The proof technique in \cite{smoller1994shock} relies on sufficient 
regularity of the solutions of elliptic (parabolic) equations. This may not hold in the presence of point sources. 
Existence results for point sources are typically obtained in the context of \emph{source-type} or \emph{very singular} solutiuons. For example, in \cite{kamin1985source} a parabolic initial value problem with polynomial non-linearity
is considered:
\begin{equation}
u_t = \Delta (u^m) - u^p, \quad x \in \mathbb{R}^d, \; t>0,
\end{equation}
where $p>1$ and a point source is in the initial condition
\begin{equation}
u(x,0) = \delta(x).
\end{equation}
Existence for the case $1<p<m+(2/d)$ is shown by approximating the point source with a sequence of smooth functions, 
while the non-existence for $p>m+(2/d)$ is established by a scaling argument. Similarly, an approximation technique 
can be used \cite{kamin1985singular} to establish existence of a solution of an elliptic equation with a point 
source in the right hand side
\begin{equation}
- \Delta u  +  u^p =  c \delta(x), \quad x \in \mathbb{R}^d,
\end{equation}
with $p < d/(d-2)$ and $c > 0$.

Note that none of the existence results mentioned above is general enough to encompass the system (\ref{eqn:usystem})
that we would like to study. Since the main focus of this work is to develop methods of solving the source identification
problem numerically, in what follows we assume for convenience that the system (\ref{eqn:usystem})--(\ref{eqn:uic}) has a 
unique solution that can be obtained as a limit 
$\boldsymbol u(\boldsymbol x, t) = \lim\limits_{q \to \infty} \boldsymbol u^q(\boldsymbol x, t)$ of an iteration
\begin{equation}
\boldsymbol u^{q+1}_t = 
\left( \boldsymbol D \boldsymbol \Delta - \boldsymbol w \cdot \boldsymbol \nabla 
+ \boldsymbol L + \boldsymbol Q(\boldsymbol u^q) \right) \boldsymbol u^{q+1}
+ \boldsymbol f, \quad q=0,1,\ldots
\label{eqn:uiter}
\end{equation}
where for each $q$ we solve the linear system (\ref{eqn:uiter}) with boundary and initial conditions 
(\ref{eqn:ubc})--(\ref{eqn:uic}) for $\boldsymbol u^{q+1}$ while keeping the previous iterate $\boldsymbol u^q$ fixed, 
starting from $\boldsymbol u^0(\boldsymbol x, t) \equiv 0$.

\subsection{Formal adjoint and source identification problem}
\label{sec:adjoint}

A straightforward way to formulate the source identification problem is to state it as an optimization problem with
PDE constraints. However, making additional assumptions on the source term like those in (\ref{eqn:fsrc}) makes it
possible to reduce the source identification problem to solving the system of non-linear algebraic equations. 
These equations arise from the formally adjoint problem. 

Let us define the inner product for vector-functions $\boldsymbol u$ and $\boldsymbol v$ by
\begin{equation}
\left< \boldsymbol u, \boldsymbol v \right>_{\Omega, T} = \int_{0}^{T} \int_{\Omega} 
\boldsymbol u(\boldsymbol x, t) \cdot \boldsymbol v(\boldsymbol x, t) d \boldsymbol x dt,
\label{eqn:inner}
\end{equation}
where $\boldsymbol u \cdot \boldsymbol v = \sum\limits_{j=1}^{n} u_j v_j$ is the inner product in $\mathbb{R}^n$.
For functions that are defined on the boundary we replace $\Omega$ in (\ref{eqn:inner}) by $\partial \Omega$, and 
when time integration is not needed we omit $T$.

To define a system formally adjoint to the non-linear system (\ref{eqn:usystem}) we observe that if the value of the 
term $\boldsymbol Q(\boldsymbol u)$ is known and fixed at the true solution $\boldsymbol u$, then (\ref{eqn:usystem}) 
is a linear system for $\boldsymbol u$. The system of equations adjoint to that linear system is given by
\begin{equation}
- \boldsymbol v_t = 
\boldsymbol D \boldsymbol \Delta \boldsymbol v 
+ \boldsymbol  w \cdot \boldsymbol \nabla \boldsymbol v 
+ \boldsymbol L^T \boldsymbol v 
+ \boldsymbol Q^T(\boldsymbol u) \boldsymbol v 
+ \boldsymbol g.
\label{eqn:vsystem}
\end{equation}
We refer to this system as a \emph{formal adjoint} to (\ref{eqn:usystem}). Hereafter we omit the term formal,
since we only use the adjoint in the above sense.

System (\ref{eqn:vsystem}) runs backwards in time from $t=T$ to $t=0$ and thus a \emph{terminal} condition for 
$\boldsymbol v(\boldsymbol x, T)$ has to be specified. Note that because the time runs backwards, the system 
is well-posed, unlike the backward parabolic system that also has a minus sign on the left, but runs forward in time.

The term $\boldsymbol g$ in (\ref{eqn:vsystem}) is chosen according to the measurement setup.
Since the source $\boldsymbol f$ in (\ref{eqn:fsrc}) is determined by many parameters $a_j$, $\boldsymbol y^j$
(and possibly also $\tau_j$), $j = 1,\ldots,N_s$, multiple measurements of $\boldsymbol u$ are needed in order to 
identify the source term. We denote by $\boldsymbol g^{(i)}$ a term corresponding to the $i^{th}$ measurement and 
by $\boldsymbol v^{(i)}$ the corresponding solution of (\ref{eqn:vsystem}) with $\boldsymbol g = \boldsymbol g^{(i)}$, 
$i=1,\ldots,N_m$ where $N_m$ is the number of measurements. A single measurement consists of measuring one component $u_{m_i}$ at location $\boldsymbol z^i$ either at a time instant $\theta_i$ or integrating over some time interval (usually the whole 
observation period $[0,T]$). This leads to $\boldsymbol g^{(i)}$ of the form
\begin{equation}
\fl
g^{(i)}_j (\boldsymbol x, t) = \delta_{j, m_i} \delta(t - \theta_i) \delta(\boldsymbol x - \boldsymbol z^{i}),
\quad j=1,\ldots,n, \quad i=1,\ldots,N_m,
\end{equation}
for the instantaneous measurement, and
\begin{equation}
g^{(i)}_j (\boldsymbol x, t) = \delta_{j, m_i} \delta(\boldsymbol x - \boldsymbol z^{i}),
\quad j=1,\ldots,n, \quad i=1,\ldots,N_m,
\end{equation}
for the measurement integrated in time. We denote the measured \emph{data} vector by
\begin{equation}
d_i = \left< \boldsymbol g^{(i)}, \boldsymbol u \right>_{\Omega, T}, \quad i=1,\ldots,N_m.
\end{equation}

Taking the inner product of (\ref{eqn:usystem}) with $\boldsymbol v$ and of (\ref{eqn:vsystem}) with $\boldsymbol u$
we can apply the divergence theorem to obtain the \emph{adjoint relation}
\begin{equation}
\left< \boldsymbol f, \boldsymbol v \right>_{\Omega, T} + c(\boldsymbol u, \boldsymbol v) = 
\left< \boldsymbol g, \boldsymbol u \right>_{\Omega, T},
\label{eqn:adjrel}
\end{equation}
where the \emph{correction term} is given by
\begin{eqnarray}
c(\boldsymbol u, \boldsymbol v) & = & - \left. \left< \boldsymbol u, \boldsymbol v \right>_{\Omega} \right|_{t=0}^{t=T}
+ \left< \boldsymbol v, \boldsymbol D \frac{\partial \boldsymbol u}{\partial \nu} \right>_{\partial \Omega, T} 
- \left< \boldsymbol u, \boldsymbol D \frac{\partial \boldsymbol v}{\partial \nu} \right>_{\partial \Omega, T} 
\label{eqn:corr} \\
& & + \left< \boldsymbol u, (\nabla \cdot \boldsymbol w) \boldsymbol v \right>_{\Omega, T}
    - \left< \boldsymbol u, (\nu \cdot \boldsymbol w) \boldsymbol v \right>_{\partial \Omega, T}. \nonumber
\end{eqnarray}
The normal derivative $\frac{\partial}{\partial \nu}$ in (\ref{eqn:corr}) is understood component-wise.
Typically one imposes the boundary and initial conditions on the adjoint solution $\boldsymbol v$ to make as many 
terms of $c(\boldsymbol u, \boldsymbol v)$ zero as possible. In particular, to take care of the $t=T$ part of the
first term in (\ref{eqn:corr}) we can set the terminal condition to $\left. \boldsymbol v \right|_{t=T} = 0$. 
The second and third terms are usually dealt with by enforcing $\boldsymbol v$ to be zero on the portion of the 
boundary where $\frac{\partial \boldsymbol u}{\partial \nu} \neq 0$ and vice versa. The fourth term typically is
zero due to the assumption of divergence free advection field $\boldsymbol w (\boldsymbol x)$. Note that if the 
advection field is divergence free than the correction term only depends on the boundary and initial conditions for 
$\boldsymbol u$ and $\boldsymbol v$ that are known a priori. 

In the examples considered below we enforce $c(\boldsymbol u, \boldsymbol v) \equiv 0$ via an appropriate choice 
of boundary conditions and advection field as described in sections \ref{sec:advection} and \ref{sec:measadj3comp}. 
Under this condition using the expression for the source (\ref{eqn:fsrc}) we can rewrite the adjoint relation 
(\ref{eqn:adjrel}) as
\begin{equation}
\sum_{k=1}^{n} \sum_{j=l_k + 1}^{l_{k+1}} a_j \int_{0}^{T} h_j(t) v_k^{(i)} (\boldsymbol y^j, t) dt = d_i,
\quad i=1,\ldots,N_m.
\label{eqn:adjdata}
\end{equation}
Note that the above system of equations is linear in source intensities $a_j$ and non-linear in the 
source spatial locations $\boldsymbol y^j$ (and also possibly temporal locations $\tau_j$). In what follows it is 
convenient to express this fact in matrix-vector form as
\begin{equation}
\boldsymbol V(\boldsymbol s) \boldsymbol a = \boldsymbol d.
\label{eqn:adjvector}
\end{equation}
Here we stack all the source intensities in the vector $\boldsymbol a$ and all source location parameters 
(including the time location parameters $\tau_j$) in vector $\boldsymbol s$ with 
$\boldsymbol s^j = (\boldsymbol y^j, \tau_j)^T$, $j=1,\ldots,N_s$ and $N_s$. If the time dependent part of the 
source term is a known indicator function, then we simply have $\boldsymbol s^j = \boldsymbol y^j$, $j=1,\ldots,N_s$.

\begin{defn}[Source identification problem]~
Given the measured data $\boldsymbol d$ taken at measurement locations $(\boldsymbol z^j, \theta_j)$, $i=1,\ldots,N_m$ 
find the source intensities $a_j$ and source location parameters $\boldsymbol s^j$, $j=1,\ldots,N_s$, 
that satisfy the adjoint relation (\ref{eqn:adjdata}). 
\end{defn}

The above definition implies that if the adjoint solutions $\boldsymbol v^{(i)}$ are known, the source identification 
problem is equivalent to solving the system of non-linear algebraic equations (\ref{eqn:adjdata}). This is indeed the 
case for the linear system, i.e. if $\boldsymbol Q(\boldsymbol u) \equiv 0$. However, in the non-linear case the adjoint 
solutions $\boldsymbol v^{(i)}$ are implicitly dependent on the forward solution $\boldsymbol u$ via the 
$\boldsymbol Q^T(\boldsymbol u)$ term in (\ref{eqn:vsystem}). The forward solution in turn depends on the source 
term $\boldsymbol f$, so there is an implicit dependency of the adjoint solution on the source, which must be resolved 
in order to solve (\ref{eqn:adjdata}). This is done using an iterative procedure that we present next.

\subsection{Forward-adjoint iteration for source identification}
\label{sec:iteration}

To obtain the source parameters $\boldsymbol a$ and $\boldsymbol s$ we need to solve the system of algebraic equations
(\ref{eqn:adjdata}), which requires the knowledge of the adjoint solutions $\boldsymbol v^{(i)}$. Adjoint solutions
satisfy a linear system (\ref{eqn:vsystem}) which includes the term $\boldsymbol Q (\boldsymbol u)$, so we must solve
the forward system (\ref{eqn:usystem}) for $\boldsymbol u$ with an unknown source $\boldsymbol f$. We propose the following
iterative procedure to solve the source identification problem that iterates over the solutions of both the forward and
adjoint problems simultaneously. 

\begin{alg}[Forward-adjoint iteration]~
\label{alg:fwdadj}
\begin{enumerate}
\item Obtain an initial guess $\boldsymbol u^0$ for the forward solution by solving a linear system
$$ \boldsymbol u^0_t = (\boldsymbol D \boldsymbol \Delta - \boldsymbol w \cdot \boldsymbol \nabla + \boldsymbol L) \boldsymbol u^0 $$
with boundary conditions (\ref{eqn:ubc}) and initial conditions (\ref{eqn:uic}).
\item[]\hskip-0.03\textwidth \textbf{For $q=1,2,\ldots$ do}
\item Solve the linear systems for the current estimate of the adjoint solutions
\begin{equation} 
- \boldsymbol v^{(i),q}_t = (\boldsymbol D \boldsymbol \Delta + \boldsymbol w \cdot \boldsymbol \nabla + \boldsymbol L^T + 
\boldsymbol Q^T(\boldsymbol u^{q-1})) \boldsymbol v^{(i),q} + \boldsymbol g^{(i)} 
\label{eqn:algadj}
\end{equation}
for $i=1,\ldots,N_m$ with the appropriate terminal and boundary conditions. 
\label{step:adj}
\item Form the matrix valued function $\boldsymbol V^q (\boldsymbol s)$ for (\ref{eqn:adjvector}) from the current
estimates of the adjoint solutions $\boldsymbol v^{(i),q}$.
\item Obtain the current estimates $\boldsymbol a^q$ and $\boldsymbol s^q$ of the source parameters
by solving iteratively the optimization problem
\begin{equation}
\mathop{\mbox{minimize}}\limits_{\boldsymbol a, \boldsymbol s} 
\| \boldsymbol V^q (\boldsymbol s) \boldsymbol a - \boldsymbol d \|_2^2
\label{eqn:optimize}
\end{equation} 
and form the current estimate of the source term $\boldsymbol f^q$.
\label{step:optimize}
\item Update the estimate for the forward solution by solving a linear system
\begin{equation}
\boldsymbol u^q_t = (\boldsymbol D \boldsymbol \Delta - \boldsymbol w \cdot \boldsymbol \nabla + \boldsymbol L + 
\boldsymbol Q(\boldsymbol u^{q-1})) \boldsymbol u^q + \boldsymbol f^q
\label{eqn:algfwd}
\end{equation}
with boundary conditions (\ref{eqn:ubc}) and initial conditions (\ref{eqn:uic}).
\label{step:fwd}
\end{enumerate}
\end{alg}

Convergence of the algorithm can be thought of in terms of both $\boldsymbol u^q$ converging to the true forward 
solution $\boldsymbol u$ and $\boldsymbol f^q$ converging to the true source term $\boldsymbol f$. While we are 
mainly interested in recovering the source term $\boldsymbol f$, convergence of one should imply convergence of 
the other and vice versa. The main idea is that (\ref{eqn:algfwd}) with an improving source estimate will behave 
like a fixed point iteration (\ref{eqn:uiter}). Convergence analysis appears to be complicated by the fact that 
iteration (\ref{eqn:algfwd}) and optimization (\ref{eqn:optimize}) are coupled. Thus, the proof of convergence 
remains to be a topic of further study.

Since the residual in the objective in (\ref{eqn:optimize}) is linear in source intensities, we can eliminate $\boldsymbol a$
from the optimization by taking the least squares solution
\begin{equation}
\boldsymbol a = \left( \boldsymbol V^T(\boldsymbol s) \boldsymbol V (\boldsymbol s) \right)^{-1} 
\boldsymbol V^T (\boldsymbol s) \boldsymbol d.
\label{eqn:leastsqa}
\end{equation}
If we substitute the above expression for $\boldsymbol a$ into (\ref{eqn:optimize}) the optimization problem can be 
rewritten as
\begin{equation}
\mathop{\mbox{maximize}}\limits_{\boldsymbol s} 
\boldsymbol d^T \boldsymbol V (\boldsymbol s)
\left( \boldsymbol V^T(\boldsymbol s) \boldsymbol V (\boldsymbol s) \right)^{-1} 
\boldsymbol V^T (\boldsymbol s) \boldsymbol d.
\label{eqn:maximize}
\end{equation}
Now the optimization objective only depends on source location parameters $\boldsymbol s$. The optimization problem 
(\ref{eqn:maximize}) is constrained by $\boldsymbol s^j \in \Omega \times [0,T]$, $j=1,\ldots,N_s$. While it is 
possible to use a derivative-based approach to solve it, here we use a simple derivative-free search procedure that 
provides good results numerically and does not require any extra work to handle the constraints. The algorithm below 
summarizes the search procedure.

\begin{alg}[Derivative-free search]~
\label{alg:derivfree}
\begin{enumerate}
\item Choose an initial guess for source location parameters $\boldsymbol s$.
\label{step:inity}
\item[]\hskip-0.03\textwidth \textbf{For $p=1,2,\ldots$ do}
\begin{enumerate}
\item[]\hskip-0.03\textwidth \textbf{For $j=1,\ldots,N_s$ do}
\begin{enumerate}
\item[(ii)] Freeze all the components $\boldsymbol s^k$ of $\boldsymbol s$ for $k \neq j$ and compute the objective
\begin{equation}
J(\boldsymbol s) = \boldsymbol d^T \boldsymbol V (\boldsymbol s)
\left( \boldsymbol V^T(\boldsymbol s) \boldsymbol V (\boldsymbol s) \right)^{-1} 
\boldsymbol V^T (\boldsymbol s) \boldsymbol d
\label{eqn:optobjj}
\end{equation}
for all possible values of $\boldsymbol s^j \in \Omega \times [0,T]$.
\item[(iii)] Update the location of the $j^{th}$ source
\begin{equation}
\boldsymbol s^j = \mathop{\mbox{argmax}}\limits_{\boldsymbol r \in \Omega \times [0,T]} 
J(\boldsymbol s^1, \ldots, \boldsymbol s^{j-1}, \boldsymbol r, \boldsymbol s^{j+1}, \ldots, \boldsymbol s^{N_s}).
\label{eqn:maxjj}
\end{equation}
\label{step:updatey}
\end{enumerate}
\item[(iv)] If for all $j=1,\ldots,N_s$ the changes in step (iii) compared to iteration $p-1$ are small then stop.
\label{step:stop}
\end{enumerate}
\end{enumerate}
\end{alg}

Algorithm \ref{alg:derivfree} has an inner-outer iteration structure. At each step of the outer iteration indexed 
by $p$ the algorithm cycles through all source locations $\boldsymbol s^j$, $j=1,\ldots,N_s$ and performs an 
exhaustive search for each of them while keeping the rest fixed. While it may seem as a computationally expensive
solution, we should note that Algorithm \ref{alg:derivfree} is just a single step in Algorithm \ref{alg:fwdadj} 
and in practice it is the cheapest step. Most of the computational time in Algorithm \ref{alg:fwdadj} is spent 
computing the adjoint solutions in step (\ref{step:adj}), so the computational cost of step (\ref{step:optimize}) 
is negligible.

Different stopping criteria can be used in step (\ref{step:stop}) of Algorithm \ref{alg:derivfree}. In practice 
since the adjoint systems (\ref{eqn:algadj}) are solved on a finite grid, one can use ``no change from iteration 
$p-1$'' as a stopping criterion in step (\ref{step:stop}). Also, the number $p$ of outer iterations of Algorithm 
\ref{alg:derivfree} can be used as a stopping criterion for the iteration indexed by $q$ of Algorithm 
\ref{alg:fwdadj}. In particular, if Algorithm \ref{alg:derivfree} terminates with $p=1$ then we can terminate 
Algorithm \ref{alg:fwdadj} as well. In practice this approach will terminate before the adjoint (\ref{eqn:algadj}) 
and forward (\ref{eqn:algfwd}) solutions fully converge, so the estimate of the source strength (\ref{eqn:leastsqa})
might be slightly inaccurate due to inaccuracy in the adjoint solutions. However, such approach gives quite accurate
estimate of the source positions $\boldsymbol s$. Moreover, this saves considerable amounts of computation, because 
the expensive step (\ref{eqn:algadj}) is not performed as many times as needed for full convergence of 
(\ref{eqn:algadj}) and (\ref{eqn:algfwd}).

Since Algorithm \ref{alg:derivfree} is used inside the iterations of Algorithm \ref{alg:fwdadj}, one can 
take as an initial guess for $\boldsymbol s$ in step (\ref{step:inity}) of Algorithm \ref{alg:derivfree} 
the estimate for $\boldsymbol s$ from iteration $q-1$ of Algorithm \ref{alg:fwdadj}. Then one has only to 
determine the initial guess for $\boldsymbol s$ at the beginning of Algorithm \ref{alg:fwdadj}. While it is 
possible to use a randomly chosen guess or a guess obtained from some prior knowledge of the source position, 
we propose a systematic way of obtaining the initial guess from the measured data $\boldsymbol d$ only. 
It is summarized below.

\begin{alg}[Initial guess for source locations]~
\label{alg:initguess}
\begin{enumerate}
\item Given the initial guess $\boldsymbol v^{0}$ from step (\ref{step:adj}) of Algorithm \ref{alg:fwdadj} with 
$q=1$, assemble the matrix $\boldsymbol V^0$ assuming that there is only one source present. In this case 
$\boldsymbol V^0$ only has one column and depends on $\boldsymbol s^1$ only. Thus the optimization objective $J$ in 
(\ref{eqn:optobjj}) also depends on $\boldsymbol s^1$ only.
\item Compute the estimate of the first source location as
\begin{equation}
\boldsymbol s^1 = \mathop{\mbox{argmax}}\limits_{\boldsymbol r \in \Omega \times [0,T]} J(\boldsymbol r).
\label{eqn:maxj1}
\end{equation}
\item[]\hskip-0.03\textwidth \textbf{For $k=2,\ldots,N_s$ do}
\item Assemble $\boldsymbol V^0$ assuming that there are $k$ sources present. Fix the locations of previously 
determined sources $\boldsymbol s^j$, $j=1,\ldots,k-1$ so that the optimization objective $J$ only depends on 
$\boldsymbol s^k$.
\item Compute the estimate of the $k^{th}$ source location as
\begin{equation}
\boldsymbol s^k = \mathop{\mbox{argmax}}\limits_{\boldsymbol r \in \Omega  \times [0,T]} 
J(\boldsymbol s^1, \ldots, \boldsymbol s^{k-1}, \boldsymbol r).
\end{equation}
\end{enumerate}
\end{alg}

Note that in the case of a single source $N_s = 1$, there is no need for an initial guess since (\ref{eqn:maxj1}) is 
the same as (\ref{eqn:maxjj}). In this case we can think of $J(\boldsymbol r)$ as an \emph{imaging functional}, which 
quantifies the likelihood of the source being located at point $\boldsymbol r \in \Omega \times [0,T]$. With noiseless 
measurements and exact knowledge of the adjoint solutions the true location of the source corresponds to the point 
where the imaging functional attains its maximum.

\subsection{Measurement placement and determining the number of sources}
\label{sec:measposnumdet}

In this section we study the question of choosing the locations at which measurements are made, which we hereafter 
refer to as measurement placement. Each source in (\ref{eqn:fsrc}) is determined by at least $d+1$ parameters, which 
are the spatial location coordinates $\boldsymbol y^j$ and intensities $a_j$, and possibly also the temporal locations 
$\tau_j$, $j=1,\ldots,N_s$.
Thus, in the simplest setting of time-independent sources we need at least $d+1$ measurements per source so that the 
non-linear system (\ref{eqn:adjvector}) is formally determined. In practice it is beneficial to have an overdetermined
system (\ref{eqn:adjvector}) since having redundant data makes source detection less sensitive to noise. Aside from 
the measurement noise there is also an issue of robustness of optimization Algorithms \ref{alg:fwdadj} and 
\ref{alg:derivfree}.
In the numerical experiments we observed that having redundant measurements also increases the robustness of 
optimization, i.e. Algorithms \ref{alg:fwdadj}--\ref{alg:initguess} are less likely to get stuck in local
minima if more measurements are added.

The problem of choosing the number and positions of measurements has two aspects to it: 
\begin{enumerate}
\item Initial placement of measurements before any data is available.
\item Adding more measurements to the existing setup based on the estimate of source locations obtained 
from the data already measured.
\end{enumerate}

\begin{figure}[t!]
\begin{center}
\includegraphics[width=0.5\textwidth]{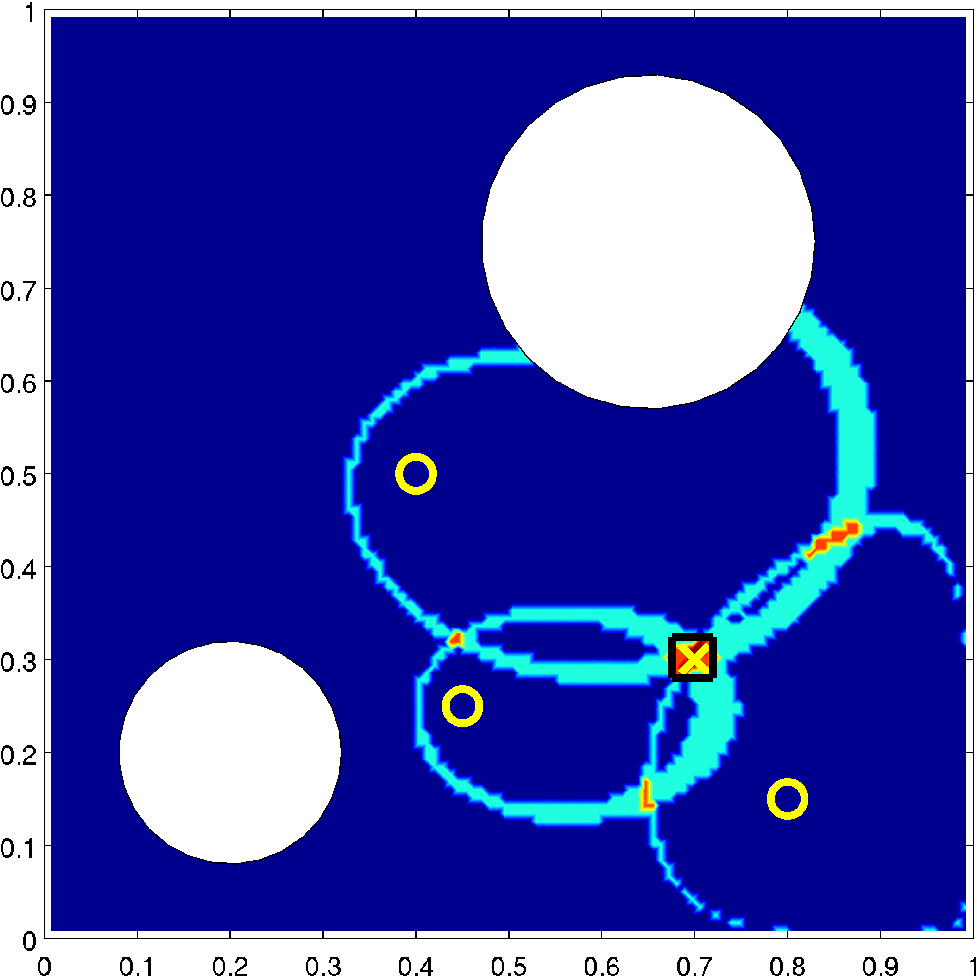}
\caption{Sum of level set indicator functions $X (\boldsymbol y)$ for the case of 
one source and three measurements. Measurement positions are yellow $\circ$, source position is yellow $\times$,
source estimate $\boldsymbol y^\star$ is black $\square$.}
\label{fig:levelset}
\end{center}
\end{figure}

To formulate a strategy of adaptive measurement placement we study how measurement positions affect the estimates
of source location in the case of a single source. Such strategy may not be optimal, because it does not take into
account the interactions between multiple sources, but it allows us to come up with a simple set of rules that can 
be applied if adding new measurements is relatively inexpensive. As we see in the numerical experiments in Section 
\ref{sec:idmultiadap} such a strategy indeed proves itself useful in the case of multiple sources.

We first consider the simplest case of identifying a single source with a known intensity. Equations 
(\ref{eqn:adjvector}) then become $\boldsymbol V(\boldsymbol y) = \boldsymbol d / \boldsymbol a$,
with both $\boldsymbol d$ and $\boldsymbol a$ scalars. Thus, the source is located at the intersection of level sets 
of the components of $\boldsymbol V(\boldsymbol y)$ corresponding to the value $\boldsymbol d / \boldsymbol a$. 
The level sets are closed curves relative to the domain $\Omega$, which is a consequence of comparison results
for parabolic equations.
There exists an analogy to the process of \emph{triangulation} in radar detection, where the corresponding curves
are circles. The analogy is exact for a linear diffusion equation in $\mathbb{R}^2$, for which the level set
curves are circles too. To illustrate this analogy numerically we consider a problem with one source and three 
measurements in two dimensions (the detailed description of the system is given in Section \ref{sec:threesystem}).
Let us introduce the indicator functions of $C$-neighborhoods of $\boldsymbol d / \boldsymbol a$ level sets
\begin{equation}
\chi_k(\boldsymbol y) = \left\{
\begin{tabular}{ll}
1, & if $|V_k(\boldsymbol y) - \boldsymbol d / \boldsymbol a| \leq C$ \\
0, & if $|V_k(\boldsymbol y) - \boldsymbol d / \boldsymbol a| > C$
\end{tabular}
\right. , \quad k=1,\ldots,3,
\end{equation}
for some $C>0$. Given the sum $X(\boldsymbol y) = \sum\limits_{k=1}^{3} \chi_k(\boldsymbol y)$ we can define the set
$S_3 = \left\{ \boldsymbol y \; | \; X(\boldsymbol y) = 3 \right\}$, then the position of the source can be estimated as
\begin{equation} 
\boldsymbol y^\star = 
\frac{\int\limits_{S_3} \boldsymbol y \; d \boldsymbol y}{\int\limits_{S_3} d \boldsymbol y}.
\end{equation}
This is shown in Figure \ref{fig:levelset} with $C=0.125$.

\begin{figure}[t!]
\begin{center}
\includegraphics[width=0.32\textwidth]{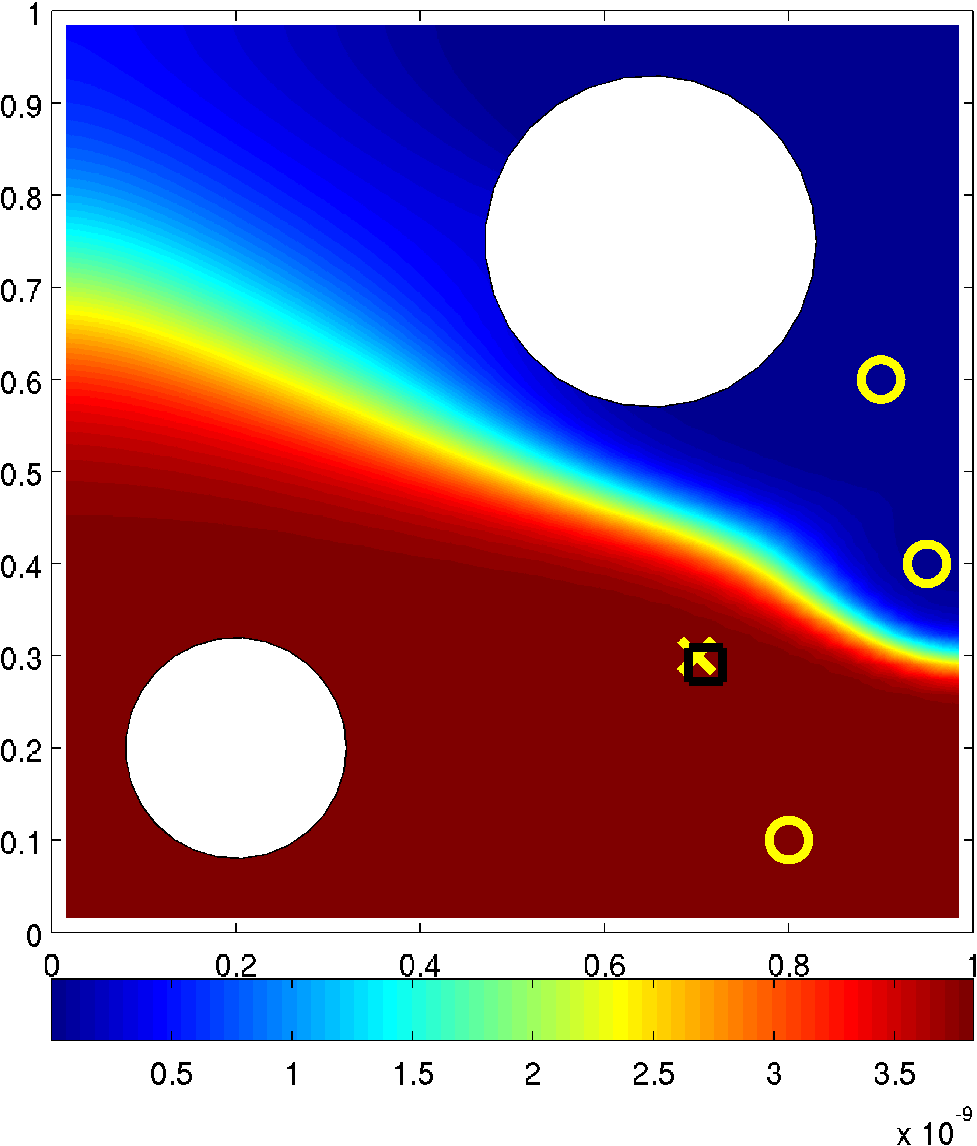}
\hskip0.01\textwidth
\includegraphics[width=0.32\textwidth]{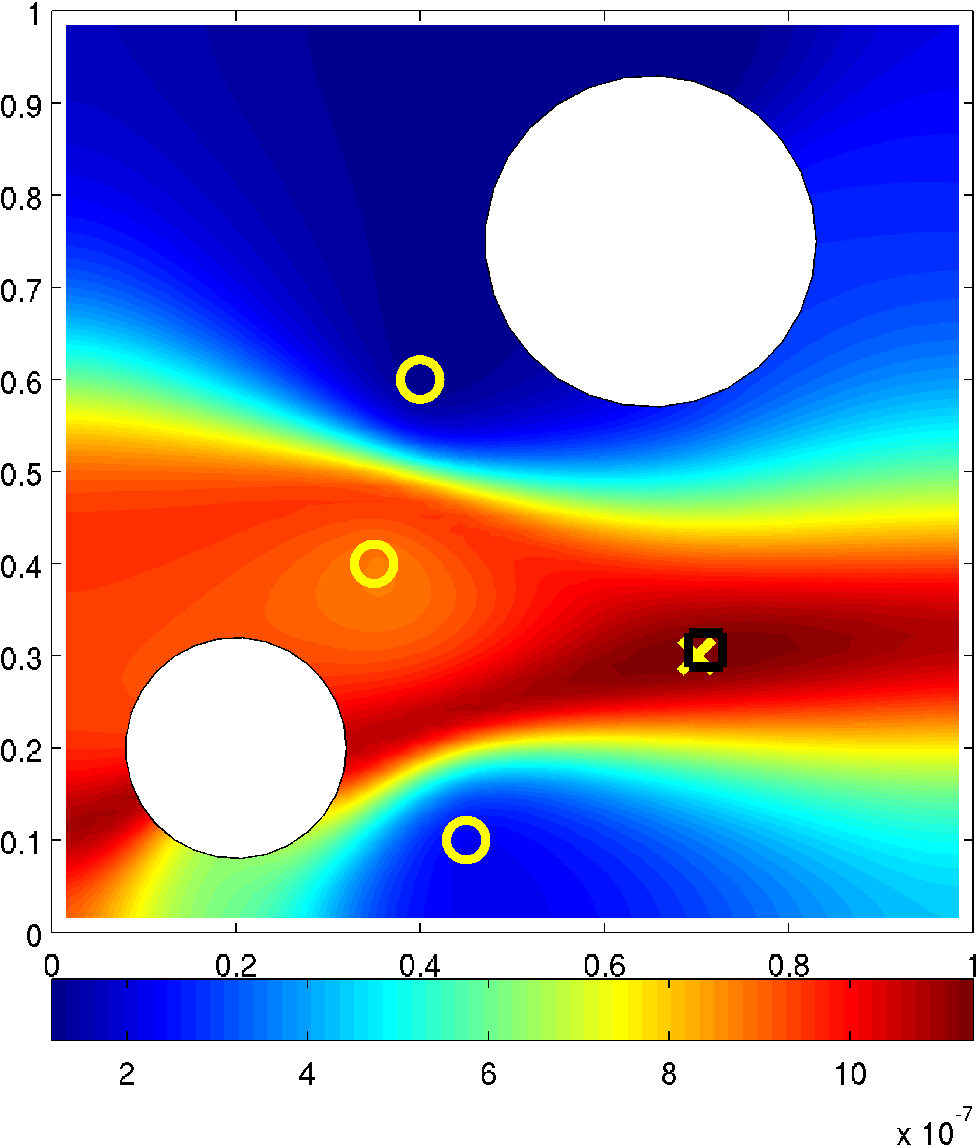}
\hskip0.01\textwidth
\includegraphics[width=0.32\textwidth]{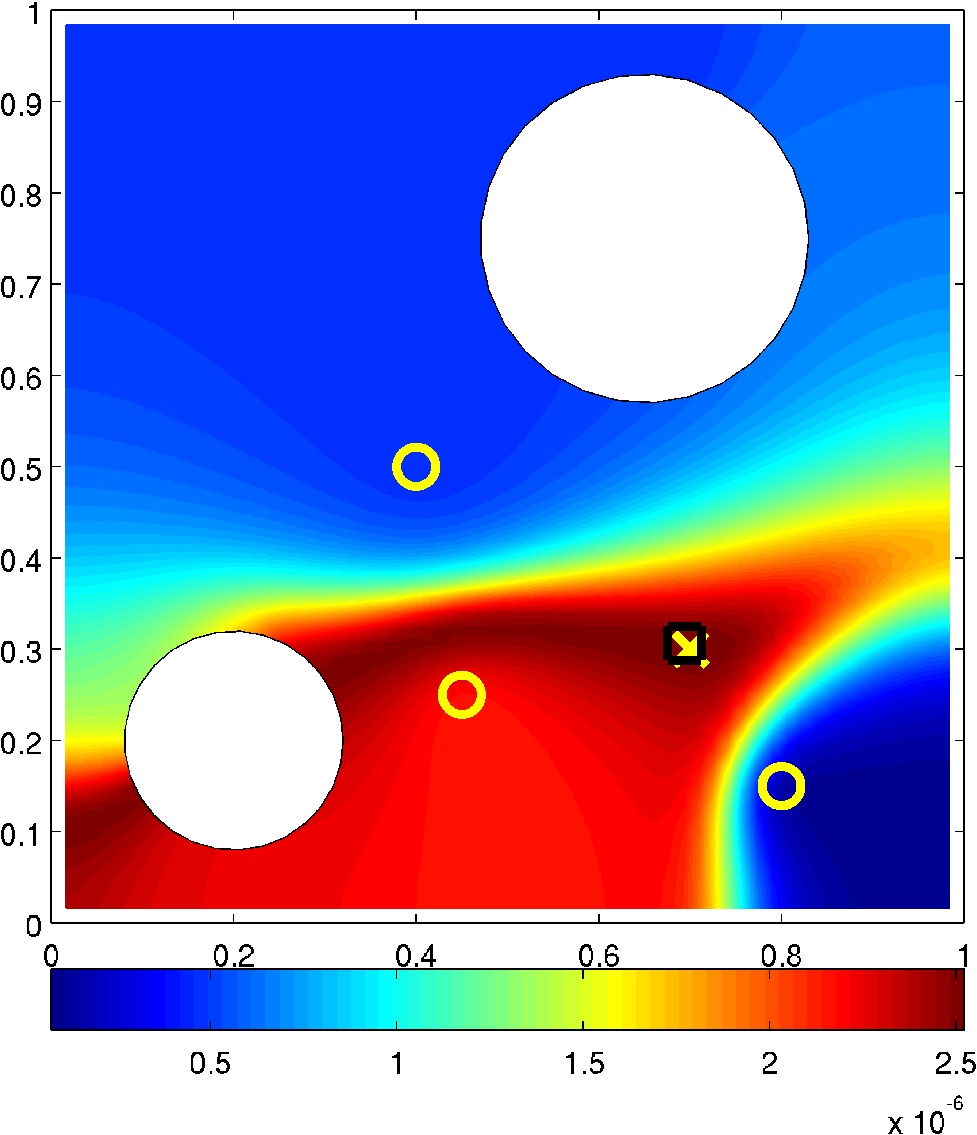}
\caption{Imaging functional $J(\boldsymbol r)$ from (\ref{eqn:maxj1}) for different measurement positions. 
Left: all measurements upwind. Middle: all measurements downwind. Right: mixed measurements (2 downwind and 1 upwind).
Measurement positions are yellow $\circ$, source position is yellow $\times$, estimated source location 
(maximum of $J(\boldsymbol r)$) is black $\square$.}
\label{fig:imagewind}
\end{center}
\end{figure}

For the case of exact data we can place the measurements anywhere in the domain and still be able to recover the  
location of the source. However, the presence of noise effectively limits the distance from the measurement to the 
source that allows a stable source identification. This is due to the fact that the magnitude of the measured
solution $\boldsymbol u$ decays quickly away from the source, and measuring weak signals is more prone to errors
than measuring strong signals. Thus, if a priori information about the source locations is not available, a 
reasonable strategy is to distribute the measurements more or less uniformly around the domain $\Omega$.

The situation is different when some prior knowledge about source locations is a available. One example is when we
would like to add new measurements adaptively based on the results of source identification with a previously chosen
smaller number of measurements. Here we assume that we can add new measurements anywhere in the domain and that
it is relatively inexpensive to do so. This leads us to a strategy of adding a few new measurements for each
previously identified source. We also assume that the sources and measurements are active for all times $t \in [0,T]$,
so only the spatial placement of the measurements is considered. For a source with unknown intensity we add
$d+1$ measurements, where we use $d=2$ dimensions for the convenience of visualization.

In order to distribute the newly added measurements around the previously estimated source locations we study how 
the distribution of measurements affects the source identification in the presence of advection in the case of a single
source. In Figure \ref{fig:imagewind} we plot the imaging functional for the three different measurement distribution
(the details of the numerical setup are given in sections \ref{sec:threesystem} and \ref{sec:numerics}). The \emph{preferred}
advection direction in Figure \ref{fig:imagewind} is from right to left, so we refer to the measurements to the right of the 
source as upwind and to the left of the source as downwind (see section \ref{sec:advection} for a definition of a 
preferred advection direction). The three possible distributions given are for all three measurements upwind, all three
measurements downwind and a mixed distribution of one measurement upwind and 2 downwind.

The plots in Figure \ref{fig:imagewind} are for the noiseless data, so the source position is recovered exactly
(up to the nearest computational grid point). However, there is a drastic difference in the behavior of the imaging
functional, which allows us to identify an optimal placement of measurements. Obviously, having all measurements upwind 
is the worst scenario. Advection propagates the plume away from the measurements and makes source identification 
difficult. This is reflected in the imaging functional having a vast plateau which implies the lack of discriminatory 
power of such functional. Ideally an imaging functional should have a single concentrated peak at the source location.
By placing all three measurements downwind the behavior of the imaging functional is much improved. The peak is now
located on a narrow ridge, thus the solution is expected to be less susceptible to noise. Finally, we observe that having
one measurement upwind can further improve the imaging functional since it allows for exclusion of a portion of the 
domain from possible source locations (the imaging functional is small around the upwind source).

Considering the above observations we propose the following procedure for adaptive measurement placement. 
\begin{alg}[Geometric adaptive measurements placement]~
\label{alg:adapmeas}
\begin{enumerate}
\item Obtain an estimate of source locations $\boldsymbol y$.
\item Choose a trust radius $\rho_T$ and a reference simplex $\boldsymbol T$ with vertices $\boldsymbol T^k$, 
$k=1,\ldots,d+1$. The orientation of the reference simplex is such that one vertex lies upwind and $d$
vertices lie downwind from its center (the center of circumscribed sphere).
\item[]\hskip-0.03\textwidth \textbf{For $j=1,\ldots,N_s$ do}  
\item Place the center of the reference simplex at $\boldsymbol y^j$.
\item[]\textbf{For $k=1,\ldots,d+1$ do} 
\begin{enumerate}
\item[(iv)] Place a new measurement in the direction of the vertex $\boldsymbol T^k$ at a distance
\begin{equation} 
\fl
\rho = \min \left( \rho_T, \kappa_\Omega \; \mbox{dist} \left( \boldsymbol y^j, \partial \Omega \right), 
\kappa_{\boldsymbol y} \; \mbox{dist} \left( \boldsymbol y^j, \left\{ \boldsymbol y^i \; | \; i \neq j \right\} \right)  
\right) \label{eqn:rho} 
\end{equation}
away from $\boldsymbol y^j$, where the constants $\kappa_\Omega, \kappa_{\boldsymbol y} \in (0,1)$ determine how close
the new measurements can be placed to the boundary and the rest of the sources respectively. 
\end{enumerate}
\item[]\textbf{For $i=1,\ldots,j-1$ do} 
\begin{enumerate}
\item[(v)] Place a new measurement on a line connecting $\boldsymbol y^j$ and $\boldsymbol y^i$.
\end{enumerate}
\end{enumerate}
\end{alg}

\begin{figure}[t!]
\begin{center}
\includegraphics[width=0.4\textwidth]{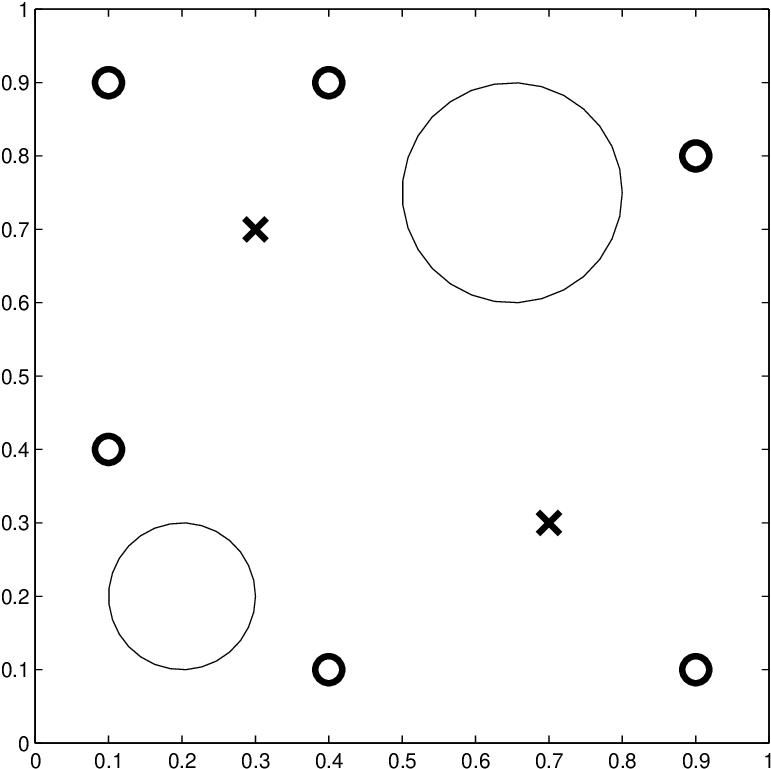} \hskip0.1\textwidth
\includegraphics[width=0.4\textwidth]{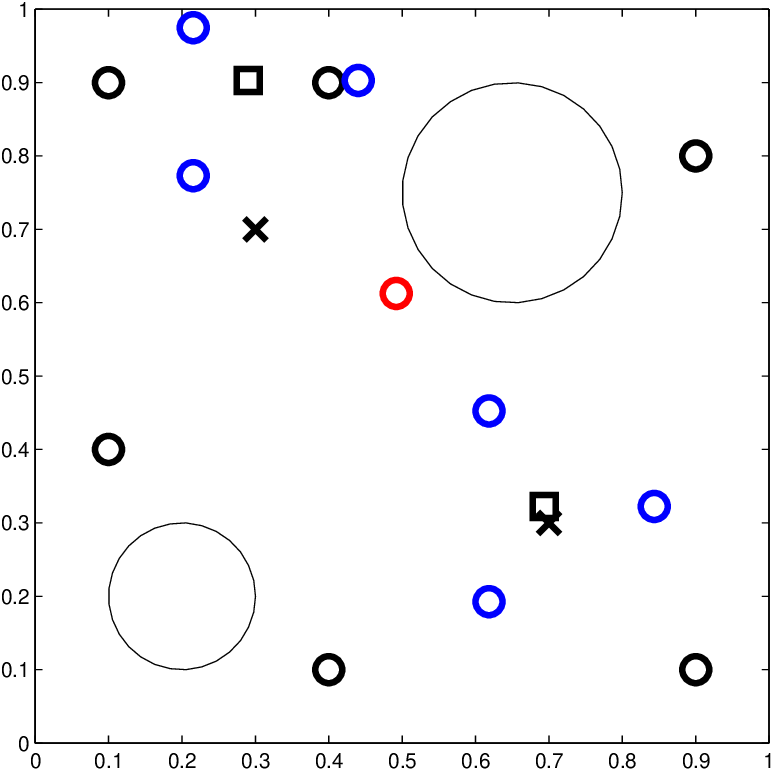}
\caption{Numerical example of performance of Algorithm \ref{alg:adapmeas}. Problem with $N_s=2$ sources (black $\times$) 
and $N_m=6$ initial measurements (black $\circ$). Left: initial measurement distribution. Right: source location estimates
(black $\square$) with noisy data ($5\%$ noise) and measurement locations added adaptively by Algorithm \ref{alg:adapmeas}
(blue and red $\circ$). Measurement locations added for better refinement in step (iv) are in blue, 
and that added in step (v) for improved separation is in red.}
\label{fig:adapmeas}
\end{center}
\end{figure}

The choice of a trust radius $\rho_T$ in step (ii) should be determined by the noise level, i.e. the distance from the
source to the measurements for which the source can be identified in a stable manner. In two dimensions the algorithm 
places three new measurements per identified source in a triangular pattern around each source so that one measurement 
is placed upwind and two downwind. Relation (\ref{eqn:rho}) ensures that the new measurements are not placed too close 
to the boundary or to other sources. This helps to separate the sources in case they are clustered together. Adding 
measurements in step (v) also helps separating clustered sources. It is possible to adjust the shape of the reference 
simplex $\boldsymbol T$ in step (v) and the positioning of the measurements in step (v) to take into account the
knowledge of the advection field. However, for simplicity in the numerical examples in Section \ref{sec:numerics} we 
use an equilateral triangle $\boldsymbol T$ in step (ii) and $\boldsymbol z = (\boldsymbol y^j + \boldsymbol y^i)/2$ 
in step (v). 

We illustrate the behavior of Algorithm \ref{alg:adapmeas} numerically in Figure \ref{fig:adapmeas}. We first obtain
the estimates of the locations of two time-independent sources based on a minimal number of measurements (six)
using Algorithm \ref{alg:fwdadj}. Algorithm \ref{alg:adapmeas} then adds seven more measurements (six in step (iv)
and one in step (v)). Note that source identification is performed with noisy data, so the initial estimate for the 
top left source is off the true location. However, after more measurements are added adaptively, both sources 
may be recovered correctly. This is shown in the numerical experiments in section \ref{sec:idmultiadap} 
(see Figure \ref{fig:srcd2}).

Note that Algorithm \ref{alg:adapmeas} is based on the idea of local refinement, i.e. the new measurements are
placed near the estimated source positions. Such approach is in agreement with the successive sampling strategy
developed in \cite{li2011heat}, which gives good results for source identification in $L_1$ setting. For each
discovered source Algorithm \ref{alg:adapmeas} adds new measurements redundantly, which may not be the most 
efficient way if each new measurement is expensive in some sense (expensive sensors, placement in remote locations, etc).
As an alternative we propose an algorithm that is based on using the level sets of adjoint solutions. Below is a
simple version of the algorithm that adds one new measurement at a time.

\begin{alg}[Level set adaptive measurements placement]~
\label{alg:levelsetmeas}
\begin{enumerate}
\item For every possible measurement location $\boldsymbol z \in \Omega$ form a corresponding term 
$\boldsymbol g^{\boldsymbol z}$ and solve the adjoint system
$$ - \boldsymbol v^{\boldsymbol z}_t = (\boldsymbol D \boldsymbol \Delta + 
\boldsymbol w \cdot \boldsymbol \nabla + \boldsymbol L^T + 
\boldsymbol Q^T(\boldsymbol u^{q})) \boldsymbol v^{\boldsymbol z} + \boldsymbol g^{\boldsymbol z}, $$
with the reaction term fixed around the last estimate $\boldsymbol u^{q}$ of the forward-adjoint iteration.
\item Select the signal level $\epsilon$ that can be measured stably.
\item Define the indicator functions of $\epsilon$-level sets
\begin{equation} 
\chi_{\boldsymbol z}^{\epsilon}(\boldsymbol x) = 
\left\{ \begin{tabular}{ll} 
$1$, & $v_k^{\boldsymbol z} (\boldsymbol x) \geq \epsilon$ \\
$0$, & $v_k^{\boldsymbol z} (\boldsymbol x) < \epsilon$ 
\end{tabular} \right., \quad \boldsymbol x \in \Omega,
\label{eqn:levelsetind}
\end{equation}
where $k$ is the index of the component of $\boldsymbol f$ that contains the (single) source.
\item Define the set
\begin{equation}
S_{\boldsymbol z}^\epsilon = \{ \boldsymbol x \in \Omega \; | \; 
\sum_{i=1}^{N_m} \chi_{\boldsymbol z_i}^\epsilon(\boldsymbol x) + \chi_{\boldsymbol z}^\epsilon(\boldsymbol x) \geq 2 \}
\label{eqn:levelset2}
\end{equation}
for every $\boldsymbol z \in \Omega$.
\item The new measurement $\boldsymbol z^\star$ is a solution of a constrained optimization problem
\begin{equation}
\boldsymbol z^\star = \mathop{\mbox{argmax}}\limits_{\mbox{s.t. } \chi_{\boldsymbol z}^\epsilon(\boldsymbol y^{q}) = 1} 
\int_{S_{\boldsymbol z}^\epsilon} d \boldsymbol x
\label{eqn:levelsetopt}
\end{equation}
\end{enumerate}
\end{alg}

Algorithm \ref{alg:levelsetmeas} is based around the idea of a region of stable identification. In the presence of noise
in the measured data we define in step (ii) of the algorithm the signal level $\epsilon$ that can be measured stably.
Thus, at each measurement location $\boldsymbol z \in \Omega$ a signal from a source located in the level set 
$\{ \boldsymbol x \in \Omega \; | \; \chi_{\boldsymbol z}^{\epsilon}(\boldsymbol x)  = 1\}$ can be stably measured.
Since a source can only be identified with multiple measurements, in the construction of $S_{\boldsymbol z}^\epsilon$
in step (iv) of the algorithm we require that two or more level sets from the measurements at existing 
($\boldsymbol z_i$) and trial ($\boldsymbol z$) locations intersect. Then, the best location for a newly added
measurement $\boldsymbol z^\star$ is the one that maximizes the coverage of $\Omega$ by such intersecting level sets,
i.e. maximizes the area of $S_{\boldsymbol z}^\epsilon$, which is the objective of (\ref{eqn:levelsetopt}). A 
reasonable constraint to have in the optimization problem (\ref{eqn:levelsetopt}) is that the estimate $\boldsymbol y^q$
from the forward-adjoint iteration with $N_m$ previously made measurements belongs to the level set of a newly added
measurement.

\begin{figure}[t!]
\begin{center}
\includegraphics[width=0.4\textwidth]{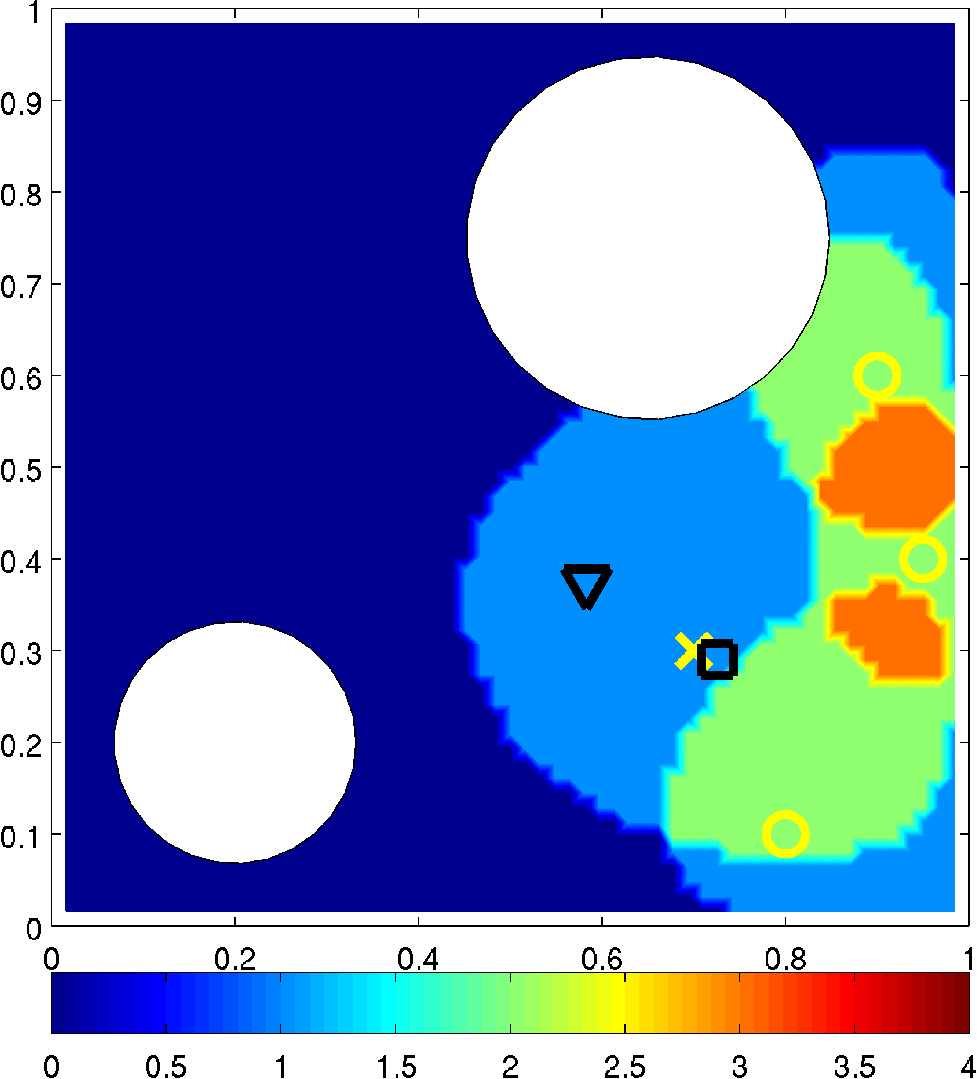} \hskip0.1\textwidth
\includegraphics[width=0.4\textwidth]{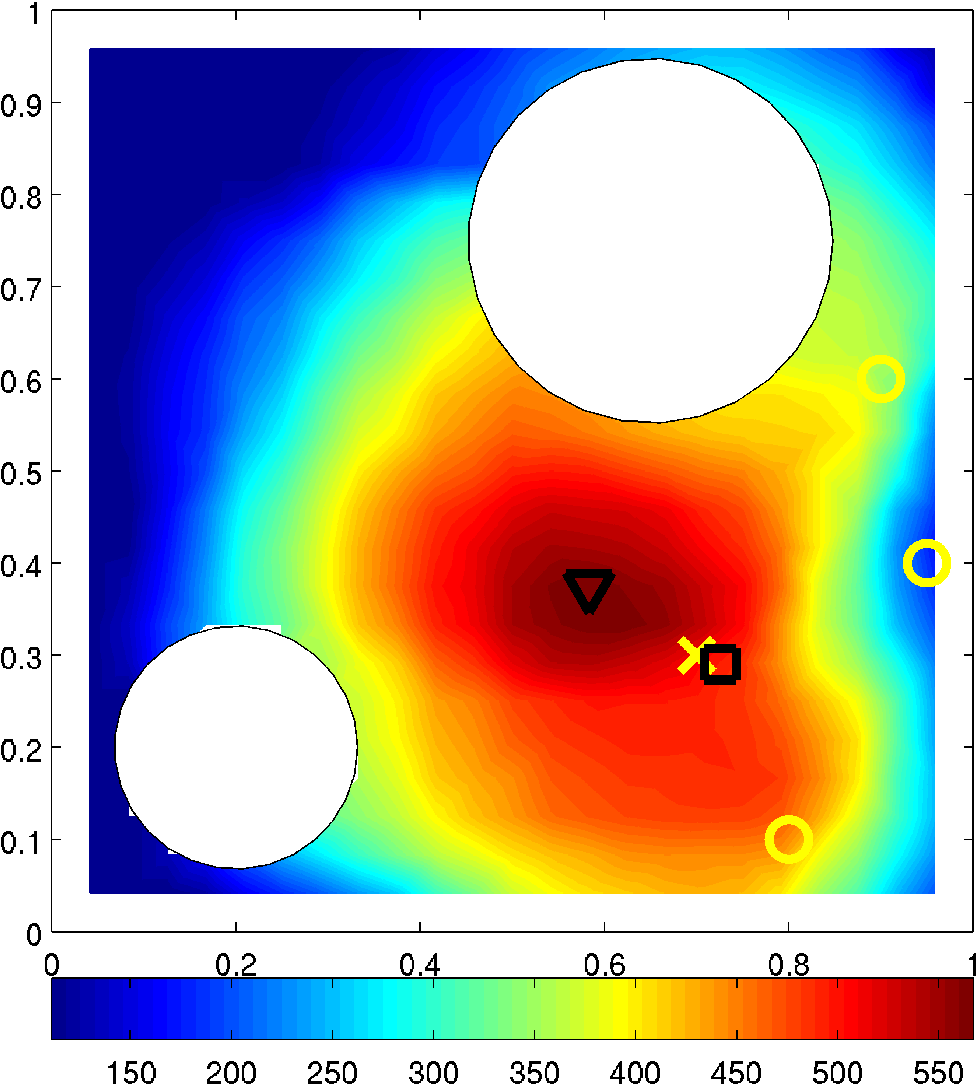}
\caption{Numerical example of performance of Algorithm \ref{alg:levelsetmeas}. Problem with a single source 
(yellow $\times$) and three measurements upwind (yellow $\circ$, same as the leftmost example in Figure \ref{fig:imagewind}).
Left: $\sum_{i=1}^{N_m} \chi_{\boldsymbol z_i}^\epsilon(\boldsymbol x) + \chi_{\boldsymbol z}^\epsilon(\boldsymbol x)$
from (\ref{eqn:levelset2}). Right: objective $\int_{S_{\boldsymbol z}^\epsilon} d \boldsymbol x$ of (\ref{eqn:levelsetopt}).
Source estimate $\boldsymbol y^{q}$ from Algoritm \ref{alg:fwdadj} is black $\square$. Newly added measurement
location $\boldsymbol z^\star$ is black $\bigtriangledown$. }
\label{fig:levelsetmeas}
\end{center}
\end{figure}

We present a numerical example of performance of Algorithm \ref{alg:levelsetmeas} in Figure \ref{fig:levelsetmeas}.
In this example we consider the case of three upwind measurements shown in the leftmost plot in Figure 
\ref{fig:imagewind}. The method performs as expected, i.e. it places the newly added measurement 
$\boldsymbol z^\star$ downwind from the source location $\boldsymbol y^q$ estimated by Algorithm 
\ref{alg:fwdadj} using three previously made measurements.
To find the solution of the optimization problem (\ref{eqn:levelsetopt}) we use a direct search over a number of
trial measurement locations. The values of the objective of (\ref{eqn:levelsetopt}) at those locations are shown 
on the right in Figure \ref{fig:levelsetmeas}. Although the objective is not convex, its behavior is quite regular,
so other optimization techniques can be employed. Note that the evaluation of the objective of (\ref{eqn:levelsetopt})
requires the computation of $\boldsymbol v^{\boldsymbol z}$ and thus the optimization problem can become quite
expensive to solve if multiple evaluations of the objective are required. This is in contrast with the geometric
approach of Algorithm \ref{alg:adapmeas} that does not require any adjoint or forward solves. Algorithm 
\ref{alg:adapmeas} is also easier to use in case when the addition of several new measurements at the same 
time is needed. These differences determine the settings in which the use of either algorithm is more beneficial. Algorithm \ref{alg:adapmeas} is advantageous when multiple measurements need to be placed and the price 
associated with deploying them is low, so some degree of redundancy can be tolerated. When placing a new 
measurement is expensive it is beneficial to use Algorithm \ref{alg:levelsetmeas} for carefully choosing an 
optimal location.

We conclude this section by considering the problem of determining an unknown number of sources, the case when $N_s$ 
is not known a priori. A procedure that appears to be both simple and reliable if to start with and estimated 
number of sources $N_s^* = 1$ and run Algorithm \ref{alg:fwdadj} repeatedly for increasing numbers 
$N_s^*=2,3,\ldots$. Note that the optimization problem (\ref{eqn:maximize}) does not impose any constraints on the
signs of components of $\boldsymbol a$. Thus, for some value of $N_s^*$ Algorithm \ref{alg:fwdadj} will compute 
a solution with $a_j = 0$ for some $j$ in the noiseless case, or in the presence of noise $a_j \leq \epsilon$ 
(this includes negative $a_j$) for some small $\epsilon$, which should be chosen based on noise level. Once this 
happens we determine the true number of sources as $N_s = N_s^* - 1$. The choice of the number of measurements in the
case of unknown $N_s$ can be done in two ways. If an upper bound $N_s < N_{s}^{\mbox{max}}$ is available one may 
set $N_m$ to the number of measurements needed to identify $N_{s}^{\mbox{max}}$ sources stably. Alternatively, one
may add the measurements adaptively using Algorithms \ref{alg:adapmeas} or \ref{alg:levelsetmeas} for each value of 
$N_s^*$. In the numerical experiments in Section \ref{sec:numunknown} we use the former approach.

\section{Three component chemical system}
\label{sec:threesystem}

In this section we consider a system that we use in the numerical experiments in Section \ref{sec:numerics}. We use a 
simplified, but somewhat realistic three component $n=3$ chemical system that models the chemical processes occurring in
the atmosphere based on Chapman's cycle \cite{sportisse2000analysis, kim1997computation}. While a realistic atmospheric 
model may contain dozens of reacting species \cite{jacobson2005fundamentals}, our simple model still captures some of 
the basic features of atmospheric models like polynomial non-linearity and stiffness.

\subsection{Forward problem}
\label{sec:fwd3comp}

The components of the system are the nitric oxide ($NO$), nitrogen dioxide ($NO_2$) and ozone ($O_3$) denoted by 
$u_1$, $u_2$ and $u_3$ respectively. We assume that nitrogen dioxide is released at source locations and the 
concentrations of nitric oxide are measured. A simplified model of chemical reactions in the system is
\begin{eqnarray}
NO + O_3 & \xrightarrow{k_1} & NO_2, \\
NO_2 & \xrightarrow{k_2} & NO + O_3,
\end{eqnarray}
with rates $k_1 = 1000$, $k_2 = 2000$. Let us introduce a scalar reaction term
\begin{equation}
r(\boldsymbol u) = k_1 u_1 u_3 - k_2 u_2,
\end{equation}
then the vector reaction term is given by
\begin{equation}
\boldsymbol R (\boldsymbol u) = 
\begin{bmatrix}
-r(\boldsymbol u) \\ 
\:\:\:r(\boldsymbol u) \\ 
-r(\boldsymbol u)
\end{bmatrix} = \boldsymbol L \boldsymbol u + \boldsymbol Q (\boldsymbol u) \boldsymbol u,
\end{equation}
where we take
\begin{equation}
\boldsymbol L = 
\begin{bmatrix}
0 & \:\:\:k_2 & 0 \\ 0 & -k_2 & 0 \\ 0 & \:\:\:k_2 & 0
\end{bmatrix}, \quad
\boldsymbol Q (\boldsymbol u) = 
\begin{bmatrix}
-k_1 u_3 & 0 & 0 \\ \:\:\:k_1 u_3 & 0 & 0 \\ 0 & 0 & -k_1 u_1
\end{bmatrix}.
\label{eqn:LQ3comp}
\end{equation}
Note that while the linear part $\boldsymbol L$ is defined uniquely, different definitions of the 
quadratic part $\boldsymbol Q$ are possible that lead to the same value of the matrix-vector product 
$\boldsymbol Q (\boldsymbol u) \boldsymbol u$ and thus the same reaction term. 

The realistic values of the diffusion constants are $\epsilon_1 = 1$, $\epsilon_2 = \epsilon_3 = 5$, which lead 
to a rather stiff system of equations due to large contrast between the diffusion constants and the reaction rates 
$k_1$, $k_2$.

While our method works in any number of spatial dimensions, in this numerical example we use $d=2$ dimensions
for the simplicity of visualization. The system is solved in the unit square with circular obstacles 
\begin{equation} 
\Omega = [0,1]^2 \setminus \left( \bigcup_{j=1}^{N_o} B_{r_j}(\boldsymbol c_j) \right),
\end{equation}
where $N_o$ is a number of obstacles. In the example below we take $N_o = 2$. Dirichlet conditions are enforced 
on the outer boundary
\begin{equation}
\left. u_1 \right|_{\partial [0,1]^2} = \left. u_2 \right|_{\partial [0,1]^2} = 0, \quad
\left. u_3 \right|_{\partial [0,1]^2} = 1,
\end{equation}
and zero Neumann conditions are enforced on the obstacle boundaries
\begin{equation}
\left. \frac{\partial u_j}{\partial \nu} \right|_{\partial B_{r_k} (\boldsymbol c_k)} = 0,
\quad j=1,\ldots,n, \quad k=1,\ldots,N_o.
\end{equation}
Constant initial conditions are used
\begin{equation}
\left. u_1 \right|_{t=0} = \left. u_2 \right|_{t=0} = 0, \quad 
\left. u_3 \right|_{t=0} = 1.
\end{equation}

We assume that at $t=0$ all sources go off and remain active for the period of time $[0, T]$. The source
term has the form
\begin{equation}
\boldsymbol f(\boldsymbol x) =
\begin{bmatrix}
0 \\ \sum\limits_{j=1}^{N_s} a_j \delta(\boldsymbol x - \boldsymbol y^j) \\ 0
\end{bmatrix},
\label{eqn:f3comp}
\end{equation}
so only the source locations $\boldsymbol y^j \in \Omega$ and the constant source intensities 
$a_j > 0$ are to be determined.

\subsection{Advection field}
\label{sec:advection}

A realistic assumption on the advection term is that there exists a preferred advection direction $\boldsymbol w_0$ 
that does not depend on time. It is also reasonable to assume that the advection vector field satisfies non-penetrating
boundary conditions on the boundaries of the obstacles
\begin{equation}
\left. (\boldsymbol w \cdot \nu) \right|_{\partial B_{r_j} (\boldsymbol c_j)} = 0, \quad j=1,\ldots,N_o.
\label{eqn:nonpenetr}
\end{equation}
Let us introduce the \emph{advection potential} $\phi$ such that
\begin{equation}
\boldsymbol w = \nabla \phi.
\end{equation}
Then the condition that the advection vector field is divergence free implies that $\phi$ must be harmonic
\begin{equation}
\Delta \phi = 0 \quad \mbox{in } \Omega,
\end{equation}
with zero Neumann boundary conditions on the obstacle boundaries
\begin{equation}
\left. \frac{\partial \phi}{\partial \nu} \right|_{\partial B_{r_j} (\boldsymbol c_j)} = 0, \quad j=1,\ldots,N_o, 
\end{equation}
and Neumann conditions enforcing the preferred direction on the outer boundaries
\begin{equation}
\left. \frac{\partial \phi}{\partial \nu} \right|_{\partial [0,1]^2} = 
\left. (\boldsymbol w_0 \cdot \nu) \right|_{\partial [0,1]^2}.
\end{equation}

Advection field used in the numerical examples below corresponds to a preferred advection direction 
$\boldsymbol w_0 = (-50, 0)$, i.e. the ``wind" blows from right to left.

\subsection{Measurements and the adjoint system}
\label{sec:measadj3comp}

For the three component chemical system we measure the component $u_1$. In the numerical results below we consider 
the cases of measurements integrated in time (sections \ref{sec:idmultiadap} and \ref{sec:numunknown}) and of instantaneous measurements (Section \ref{sec:timesrcid}). The term $\boldsymbol g$ in the adjoint system (\ref{eqn:vsystem}) for the $i^{th}$ measurement takes the form
\begin{equation}
\fl
\boldsymbol g^{(i)}(\boldsymbol x) = 
\begin{bmatrix}
\delta(\boldsymbol x - \boldsymbol z^i) \\ 0 \\ 0
\end{bmatrix}, \mbox{ or} \quad
\boldsymbol g^{(i)}(\boldsymbol x) = 
\begin{bmatrix}
\delta(t-\theta_i) \delta(\boldsymbol x - \boldsymbol z^i) \\ 0 \\ 0
\end{bmatrix},
\quad i=1,\ldots,N_m.
\end{equation}
for integrated or instantaneous measurements respectively.

While the initial and boundary conditions for the forward system typically come from the physical problem, we have a 
freedom of choosing the terminal and boundary conditions for the adjoint system, so that the adjoint relation 
(\ref{eqn:adjrel}) is as simple as possible. In particular, we would like the correction term (\ref{eqn:corr}) to be 
zero. We enforce  zero Dirichlet conditions on the outer boundary and zero Neumann conditions on the obstacle's 
boundaries for all three components of the adjoint solution $\boldsymbol v$. We also use zero terminal condition 
$\boldsymbol v(\boldsymbol x, T) = 0$, $\boldsymbol x \in \Omega$. 

From the expression (\ref{eqn:LQ3comp}) for $\boldsymbol L$ and $\boldsymbol Q(\boldsymbol u)$ we obtain the equation 
for the third component of the adjoint solution
\begin{equation}
- v_{3,t} = \left( \epsilon_3 \Delta - \boldsymbol w \cdot \nabla - k_1 u_1 \right) v_3.
\end{equation}
Combined with the terminal and boundary conditions we immediately see that 
\begin{equation}
v_3(\boldsymbol x, t) \equiv 0, \quad \boldsymbol x \in \Omega, \quad t \in [0,T].
\end{equation}
This is enough to make the correction term (\ref{eqn:corr}) zero. Indeed, the only component of $\boldsymbol u$ and 
$\boldsymbol v$ that has non-zero initial (terminal) or boundary conditions is $u_3$. Since $v_3$ is identically 
zero it neutralizes non-zero initial and boundary conditions for $u_3$ in the first three terms of the 
$c(\boldsymbol u, \boldsymbol v)$. There is no contribution from the other components of $\boldsymbol u$, and 
$\boldsymbol v$ to the first three terms, so they are identically zero. The fourth term is zero since we use 
a divergence free advection field $\boldsymbol w$. Finally, the fifth term on the outer boundary is taken
care of because $\left. \boldsymbol v \right|_{\partial [0,1]^2} = 0$. On the boundaries of the obstacles it
is zero since $\boldsymbol w$ satisfies the non-penetrating conditions (\ref{eqn:nonpenetr}) there.

Once we establish that $c(\boldsymbol u, \boldsymbol v)$ is zero we can write the components of the system of 
equations (\ref{eqn:adjvector}) arising from the adjoint relation
\begin{eqnarray} 
V_{ik} = \int_{0}^{T} v_2^{(i)} (\boldsymbol y^k, t) dt, & \mbox{or} & V_{ik} = v_2^{(i)} (\boldsymbol y^k, \tau_k), \\
d_i = \int_{0}^{T} u_1 (\boldsymbol z^i, t) dt, & \mbox{or} & d_i = u_1 (\boldsymbol z^i, \theta_i) \label{eqn:dataint},
\end{eqnarray}
for integrated or instantaneous measurements respectively, where $i = 1,\ldots,N_m$, $k = 1,\ldots,N_s$ and
\begin{equation}
\boldsymbol a = \left( a_1, \ldots a_{N_s} \right)^T,
\end{equation}
is the same for both cases.

\section{Numerical results}
\label{sec:numerics}

We implement our method of source identification and provide the results of the numerical experiments below. The first
two sets of experiments use time integrated measurements as described in Section \ref{sec:measadj3comp}. In these experiments 
we identify time-independent sources in the cases where the number of sources itself is known (Section \ref{sec:idmultiadap}) 
or unknown (Section \ref{sec:numunknown}). In Section \ref{sec:idmultiadap} we also study adaptive positioning of 
measurements and its influence on source identification. Finally, in Section \ref{sec:timesrcid} we provide the numerical 
results for identification of a time dependent source from instantaneous measurements. Results from both one and two 
dimensional settings are presented.

\subsection{Linear parabolic solver}
\label{sec:solver}

We solve the linear parabolic systems for the forward and adjoint iterations using the following numerical schemes. 
The spatial part is discretized with finite differences on a uniform Cartesian grid. The two-dimensional Laplace operator 
in the diffusion term is discretized using the standard five-point stencil. The advection term is discretized using a central
difference scheme. The reason for using the central difference scheme for the advection term is that we can use 
the same discretization for the forward and adjoint problem, for which the direction of advection is reversed. 
Note that such discretization may become inaccurate if the advection dominates other terms. While there exist more
sophisticated and accurate numerical schemes for the solution of advection-diffusion equations, the focus of this
work is not the numerical solution of the forward problem. The numerical scheme described in this section appears 
to be sufficiently accurate for source identification in a three component system described in Section 
\ref{sec:threesystem}.

To obtain the solution in time we use an exponential integrator. After discretizing in space we need to solve the 
system of ODEs for the forward and adjoint problems of the following form
\begin{equation}
\boldsymbol \xi_t = \boldsymbol E(t) \boldsymbol \xi + \boldsymbol \zeta(t).
\end{equation}
The dependency of the matrix $\boldsymbol E$ on time is due to the fact that the reaction term $\boldsymbol Q$ 
depends on the forward solution that is a function of time. If we denote the $k^{th}$ time step by $t_k$ and
the size of the step is $h_k = t_{k+1} - t_k$, then the approximate solution at time step $k+1$ is given by
\begin{equation}
\fl
\boldsymbol \xi^{(k+1)} = \mbox{exp} \left( \boldsymbol E^{(k)} h_k \right)
\left( \left( \boldsymbol E^{(k)} \right)^{-1} \boldsymbol \zeta^{(k)} + \boldsymbol \xi^{(k)} \right) 
- \left( \boldsymbol E^{(k)} \right)^{-1} \boldsymbol \zeta^{(k)},
\label{eqn:timeexpint}
\end{equation}
where $\boldsymbol \xi^{(k)} \approx \boldsymbol \xi(t_k)$, $\boldsymbol E^{(k)} = \boldsymbol E(t_k)$ and
$\boldsymbol \zeta^{(k)} = \boldsymbol \zeta(t_k)$. While each step of this method is more computationally 
expensive than that of traditional time stepping methods, it is more accurate allowing us to use a small number
of time steps. Note that (\ref{eqn:timeexpint}) requires evaluation of matrix-vector products with matrix exponentials.
We evaluate these products using an efficient algorithm \cite{al2011computing}.

In order to avoid committing an inverse crime \cite{colton1998inverse} we use different grid and time steps for 
the forward problem data simulation and for the solution of the source identification problem with Algorithm 
\ref{alg:fwdadj}. We simulate the data on a finer grid with $80$ grid nodes in both $x$ and $y$ directions 
and $30$ time steps. In the case of time independent source and integrated measurements we perform source 
identification on a grid with $63$ grid nodes in both $x$ and $y$ directions and $19$ time steps. 

Note that even without adding artificially generated noise to the data, using different (and relatively coarse) grids 
for the data simulation and source identification is equivalent to having some systematic error in the measurements. 
This poses an issue in the case of time dependent source and instantaneous measurements, since this case is more sensitive 
to the noise level in the data. In this case we use a finer grid for source identification, namely with $73$ nodes 
in both directions. Another modification to the solver required in this case is the use of non-uniform time stepping. 
In order to properly resolve the singularity of the sources around times $\tau_k$, $k = 1,\ldots,N_s$ in the
forward problem and around $\theta_j$, $j=1,\ldots,N_m$ in the adjoint problems, we refine the time stepping locally.

\subsection{Noise model}
\label{sec:noise}

We provide below the results of the numerical experiments for identifying sources from noisy measurements. 
Single source identification with noiseless measurements can be found in Figure \ref{fig:imagewind}. 
In this section we use a simple noise model with multiplicative normally distributed noise. Such model
while being easy to implement captures a realistic assumption that the noise level can be viewed as constant
relative to the strength of the signal.

If we denote the simulated data vector by $\boldsymbol d$, then the noisy data $\boldsymbol d^*$ is given by
\begin{equation}
\boldsymbol d^* = (\boldsymbol I + \sigma \boldsymbol N) \boldsymbol d, \quad
\boldsymbol N = \mbox{diag} \left( X_1, \ldots, X_{N_m} \right),
\label{eqn:noise}
\end{equation}
where $\sigma$ is a scaling term and $X_j$, $j=1,\ldots,N_m$ are independent normally distributed random variables 
with zero mean and unit standard deviation. All results are presented for one particular realization of noise, although 
for different realizations the results remain similar, which indicates that the source identification Algorithm 
\ref{alg:fwdadj} is relatively stable.

In the case of time-independent sources and integrated measurements we take the scaling factor $\sigma=0.05$ corresponding 
to $5\%$ relative noise level. Note that according to (\ref{eqn:noise}) the noise is added to the data after the integration 
in (\ref{eqn:dataint}). Adding the noise to $u_1$ before the integration in (\ref{eqn:dataint}) would make it 
easier for Algorithm \ref{alg:fwdadj} to determine the source, since integration in (\ref{eqn:dataint}) would act 
as a noise canceling filter. In order to stress test our method we add the noise after the integration instead. 
As a measure of error in the solution we use a relative location error given by
\begin{equation}
E = \frac{1}{l N_s} \sum_{j=1}^{N_s} \| \boldsymbol y^j - \widehat{\boldsymbol y}^j \|,
\end{equation}
where $\boldsymbol y^j$ are the estimates and $\widehat{\boldsymbol y}^j$ are the true source locations. 
The characteristic length $l$ is set here to $l=1$ since our domain is the unit square with obstacles. 

The case of time dependent source and instantaneous measurements is more difficult. Accordingly, we reduce the 
noise level to $\sigma=0.01$ for the numerical experiments in Section \ref{sec:timesrcid}.

\subsection{Identifying multiple sources with adaptive measurement placement}
\label{sec:idmultiadap}

\begin{figure}[t!]
\begin{center}
\includegraphics[width=0.4\textwidth]{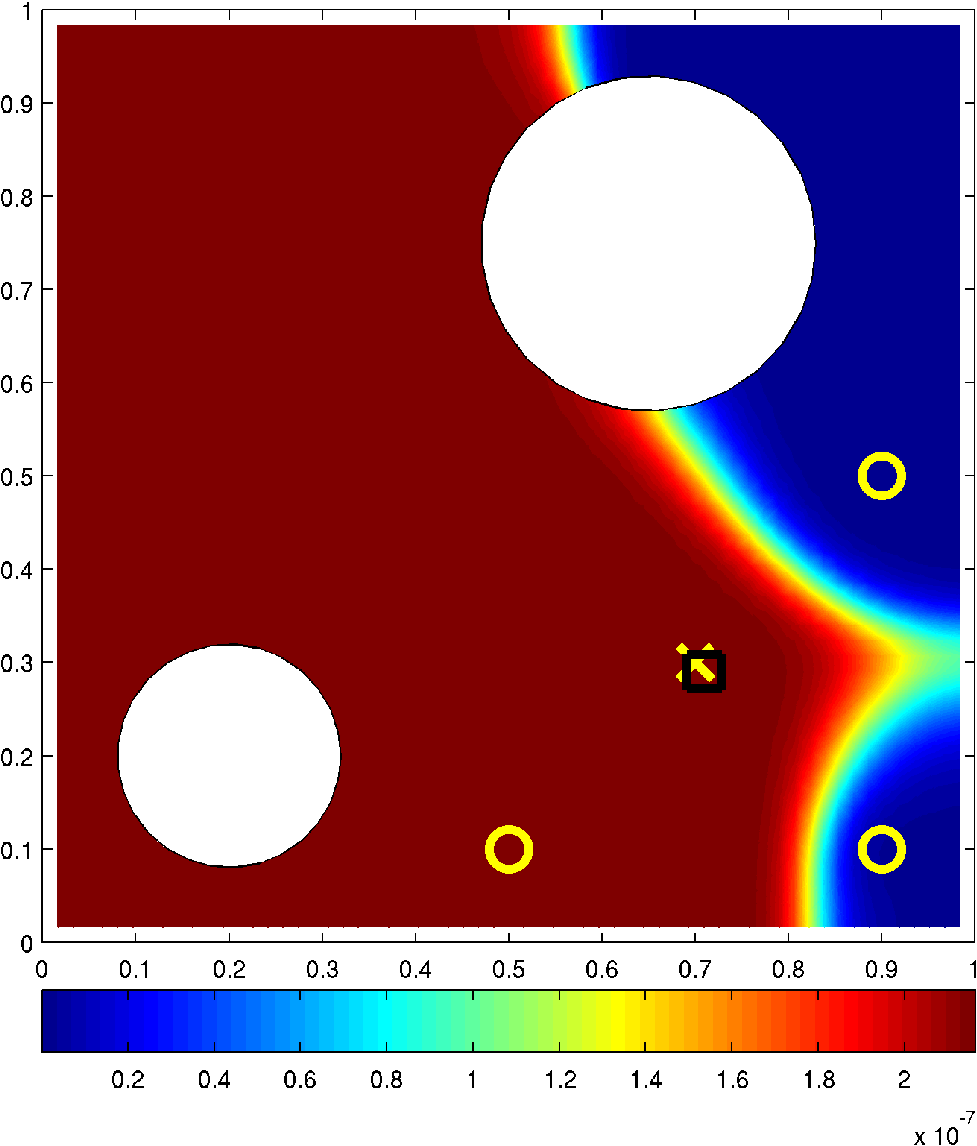} \hskip0.1\textwidth
\includegraphics[width=0.4\textwidth]{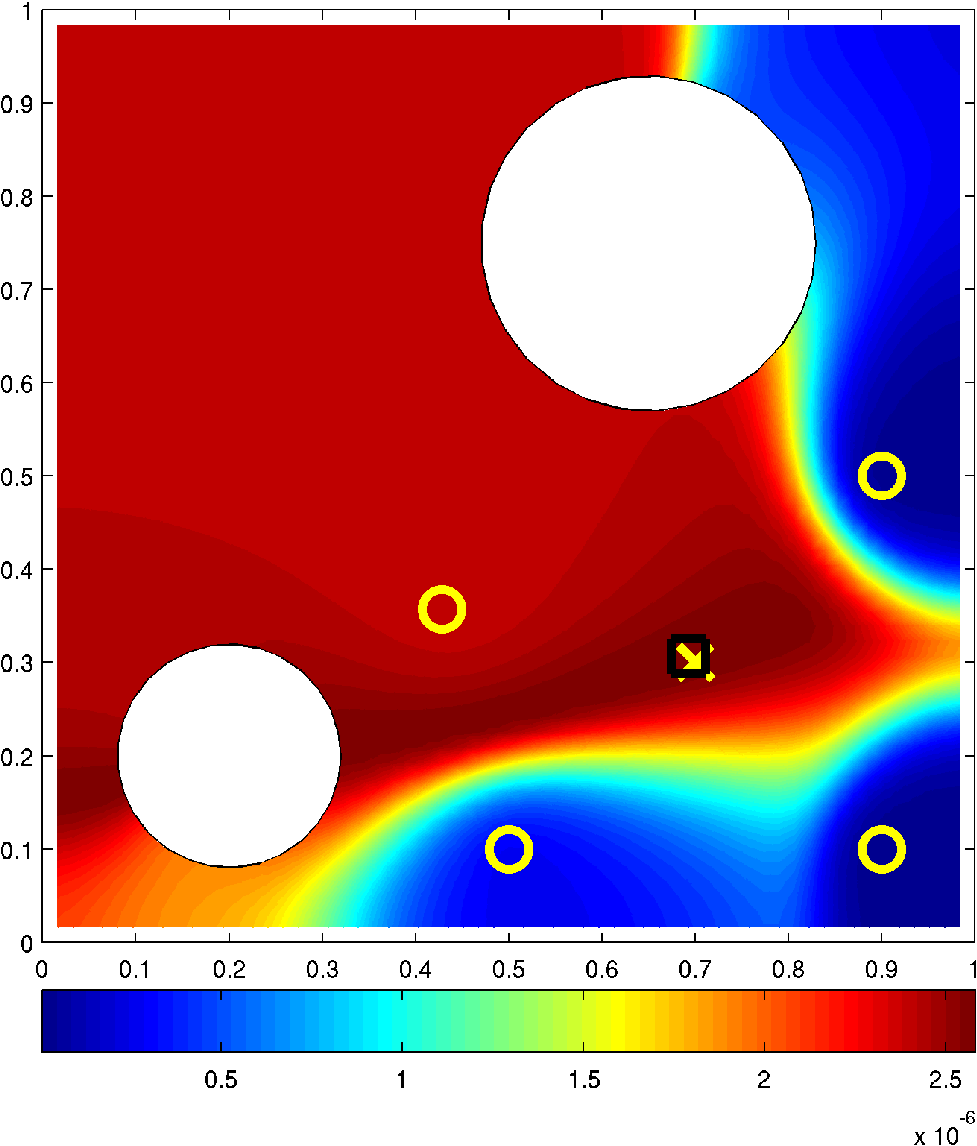}
\caption{Imaging functional $J(\boldsymbol r)$ for all $\boldsymbol r \in \Omega$ given by the final 
iteration of Algorithm \ref{alg:fwdadj}. Right: initial run with $N_m=3$ measurements. Left: subsequent run with 
an extra measurement added adaptively by Algorithm \ref{alg:levelsetmeas}.
True source location is yellow $\times$, measurement locations are yellow $\circ$,
estimated source position $\boldsymbol y$ - maximum of the imaging functional is black $\square$.
}
\label{fig:srcd1}
\end{center}
\end{figure}

\begin{figure}[t!]
\begin{center}
\includegraphics[width=0.4\textwidth]{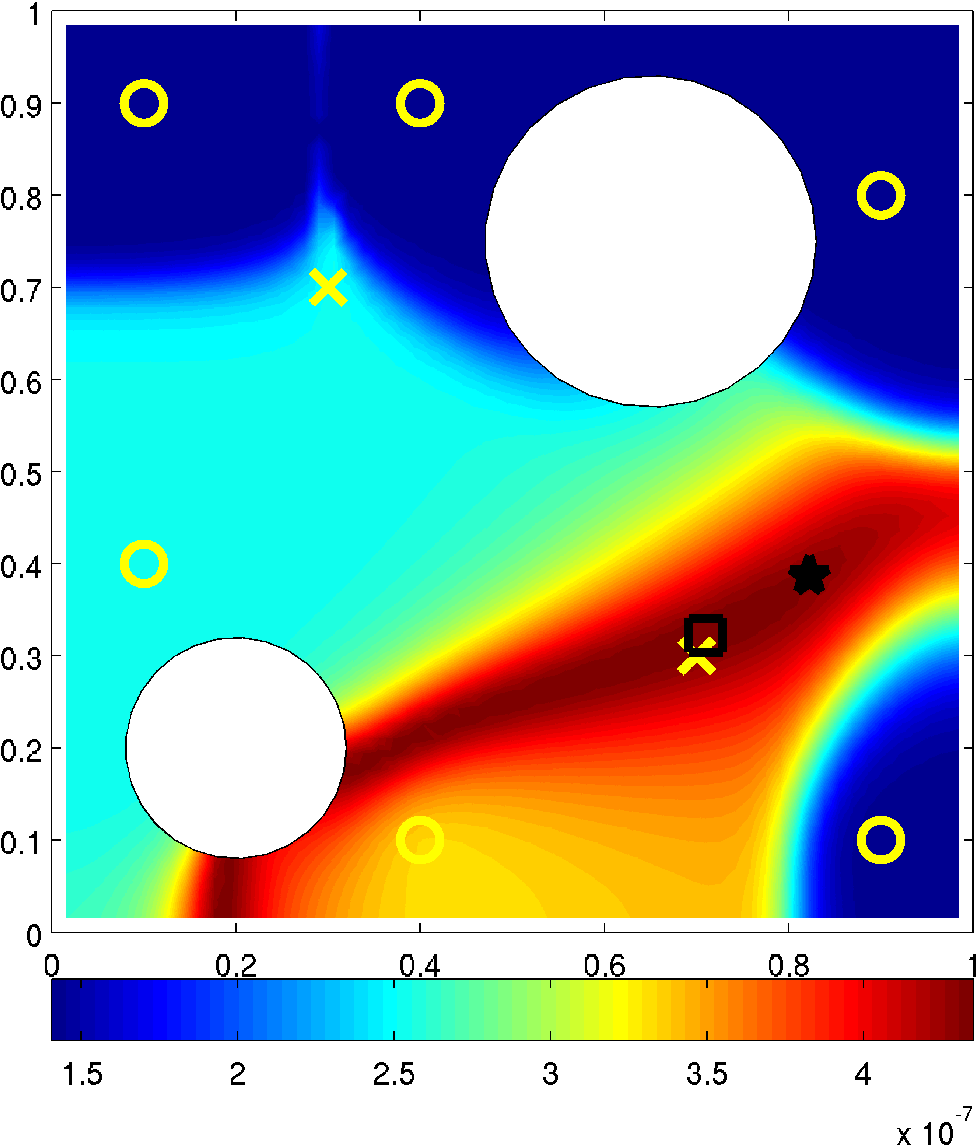} \hskip0.1\textwidth
\includegraphics[width=0.4\textwidth]{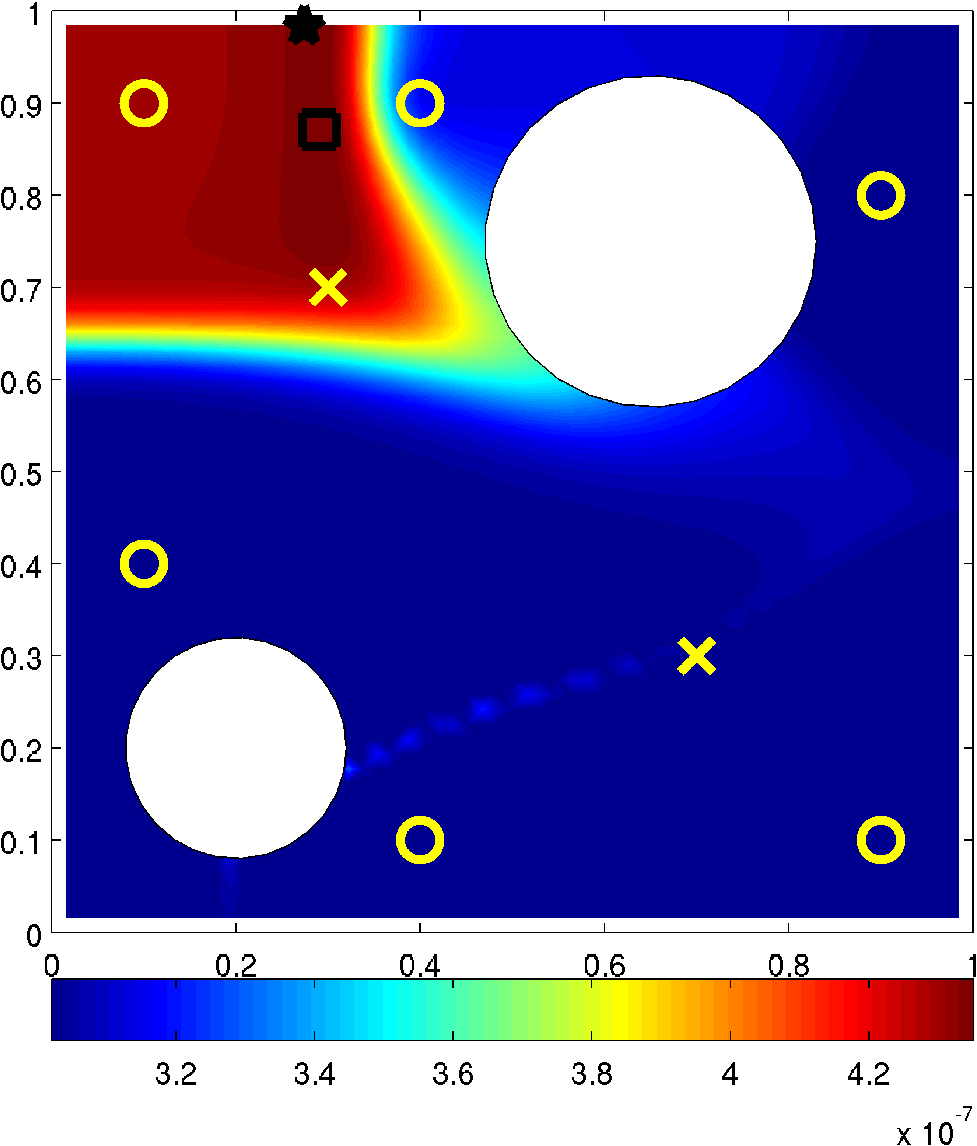}
\includegraphics[width=0.4\textwidth]{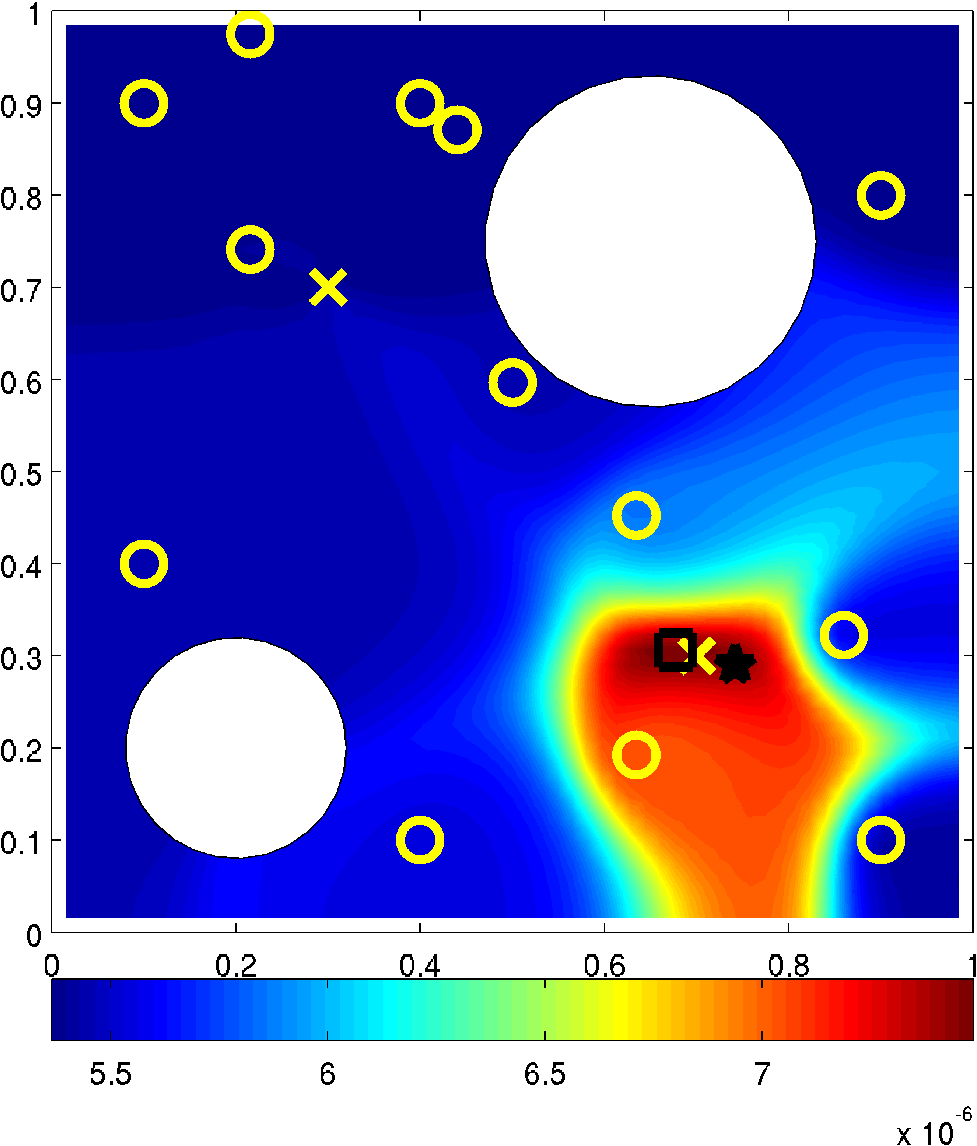} \hskip0.1\textwidth
\includegraphics[width=0.4\textwidth]{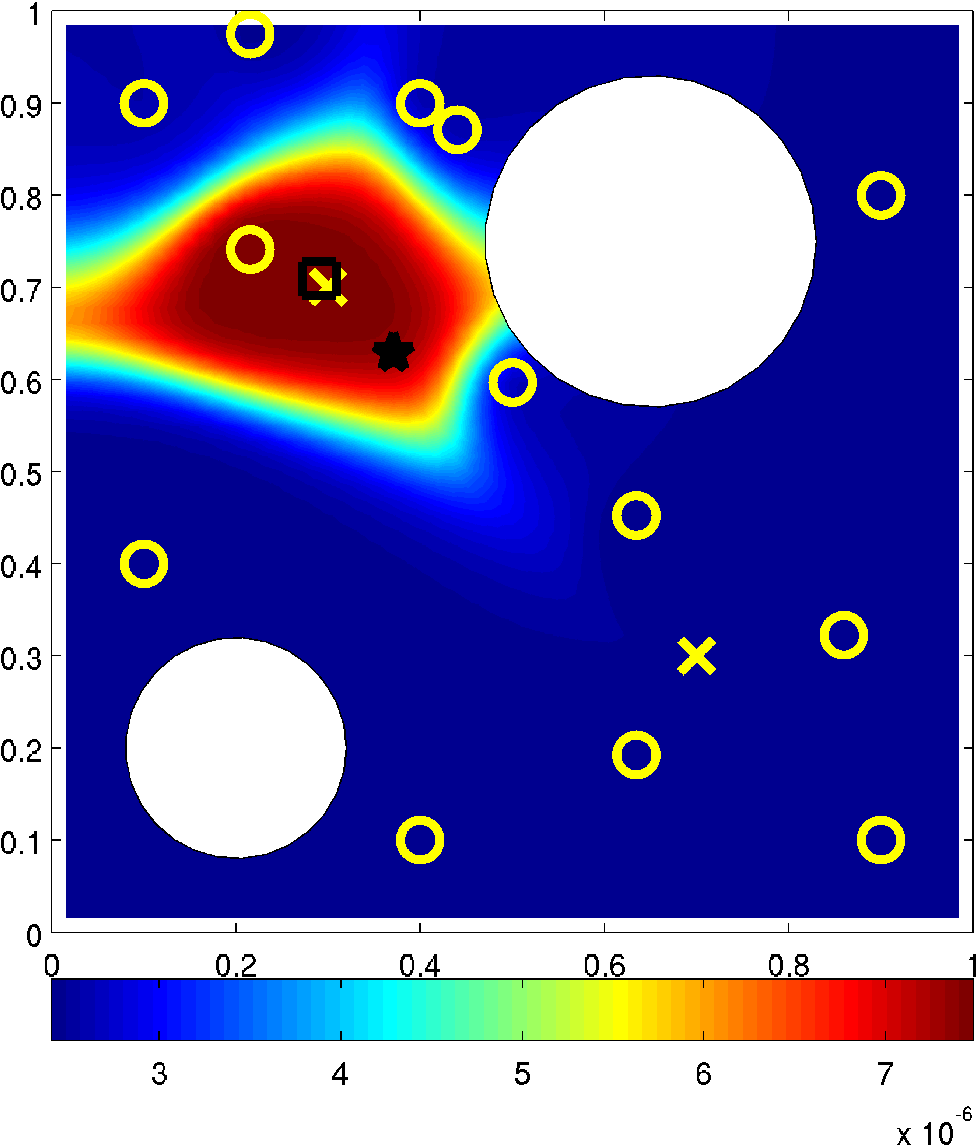}
\caption{Imaging functionals $J(\boldsymbol r, \boldsymbol y^2)$ (left column) and 
$J(\boldsymbol y^1, \boldsymbol r)$ (right column) from (\ref{eqn:maxjj}) for all $\boldsymbol r \in \Omega$ 
evaluated at the final value of $(\boldsymbol y^1, \boldsymbol y^2)$ given by Algorithm \ref{alg:fwdadj}.
Top row: initial run with $N_m=6$ measurements. Bottom row: subsequent run with measurements added adaptively
by Algorithm \ref{alg:adapmeas} ($N_m = 13$). True source locations are yellow $\times$, measurement locations 
are yellow $\circ$, estimated source position $(\boldsymbol y^1, \boldsymbol y^2)$ - maximum of the imaging 
functional is black $\square$, initial guess from Algorithm \ref{alg:initguess} is black $\star$.
}
\label{fig:srcd2}
\end{center}
\end{figure}

\begin{table}
\caption{\label{tab:srcd2}True and reconstructed source intensities $a_j$ and locations $\boldsymbol y^j$,
and relative location errors $E$ for the case of two sources ($5\%$ noise in the data).}
\begin{indented}
\item[]\begin{tabular}{@{}cccccc}
\br
Case & $a_1$ & $a_2$ & $\boldsymbol y^1$ & $\boldsymbol y^2$ & $E$ \\
\mr
True       & 10.00 & 7.00 & (0.70, 0.30) & (0.30, 0.70) & -- \\
$N_m = 6$  & 13.37 & 2.27 & (0.70, 0.32) & (0.29, 0.87) & 0.09 \\
$N_m = 13$ & 9.62 & 6.44 & (0.67, 0.30) & (0.29, 0.70) & 0.02 \\
\br
\end{tabular}
\end{indented}
\end{table}

\begin{figure}[t!]
\begin{center}
\includegraphics[width=0.31\textwidth]{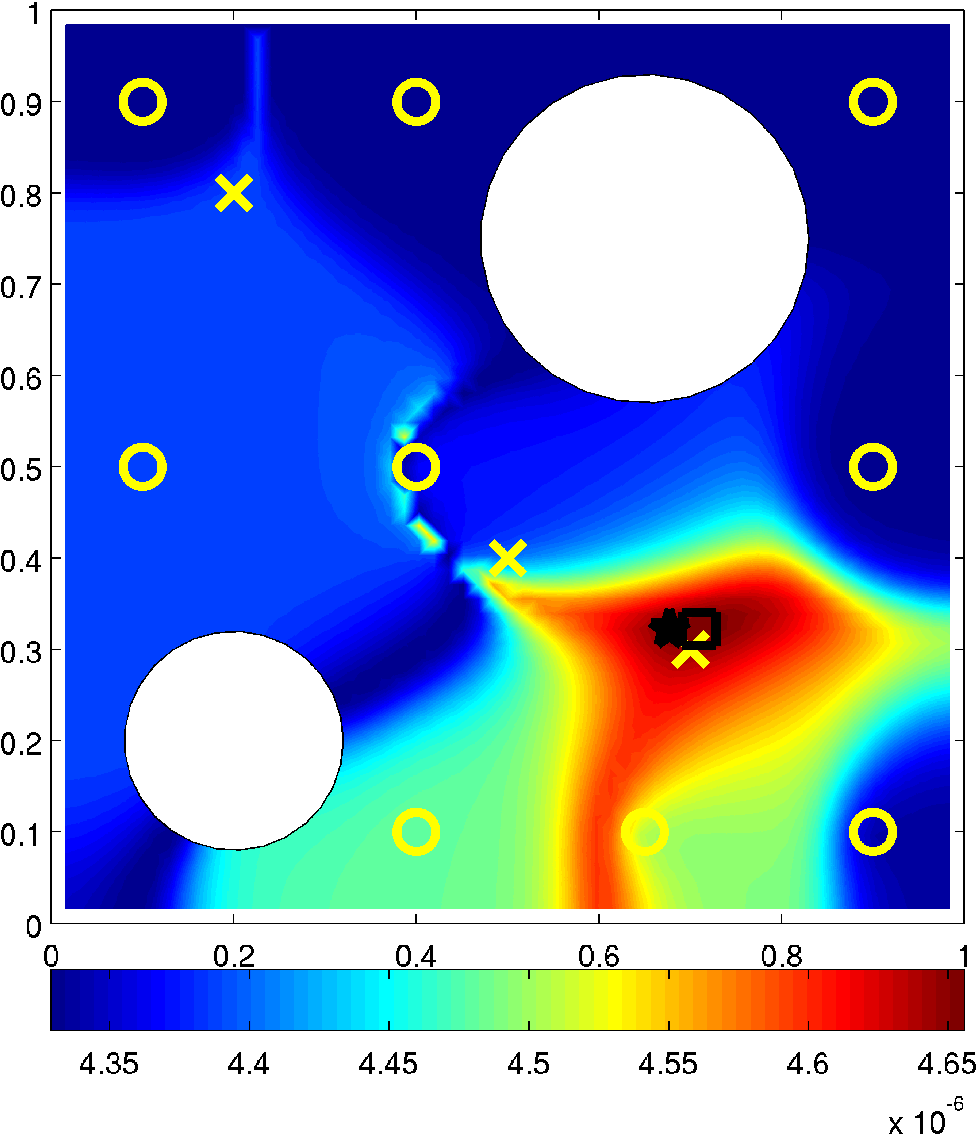} \hskip0.02\textwidth
\includegraphics[width=0.31\textwidth]{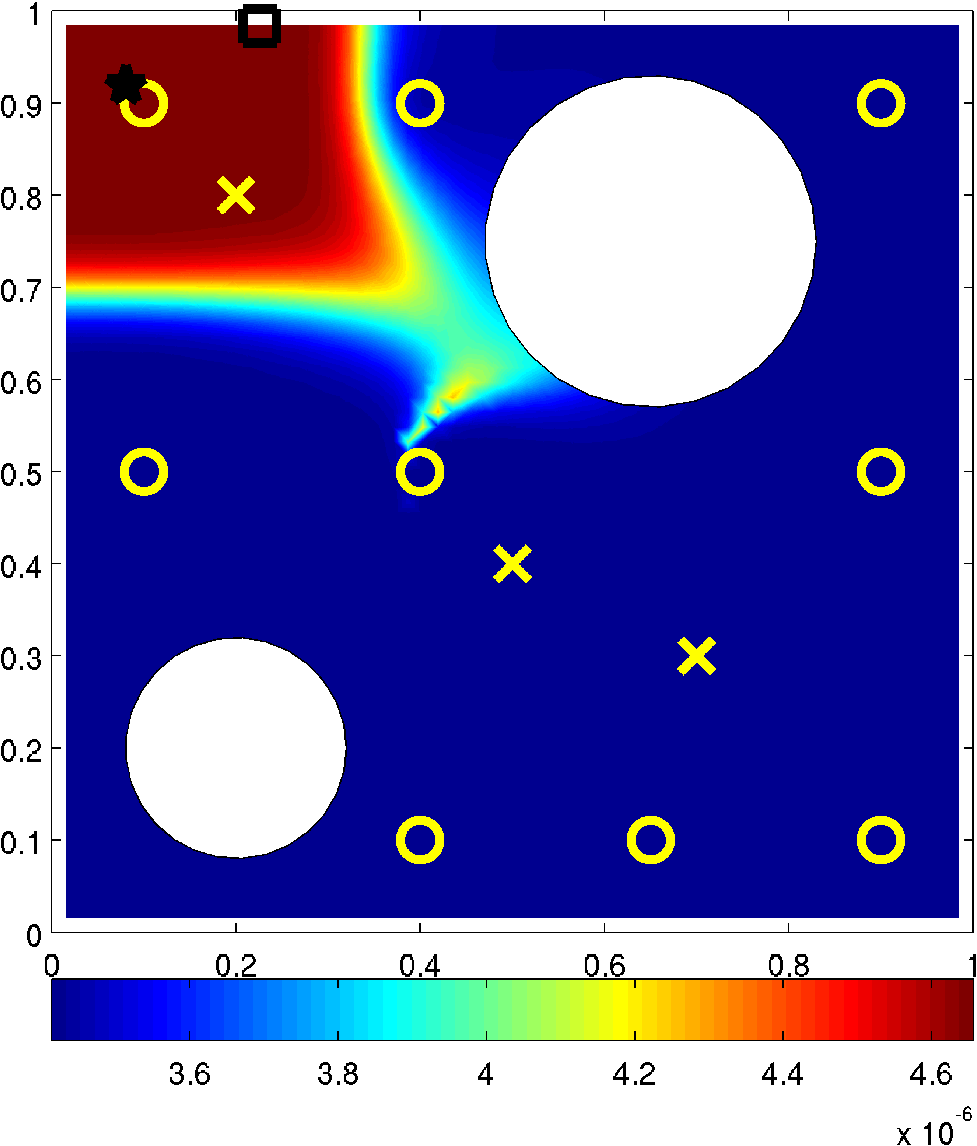} \hskip0.02\textwidth
\includegraphics[width=0.31\textwidth]{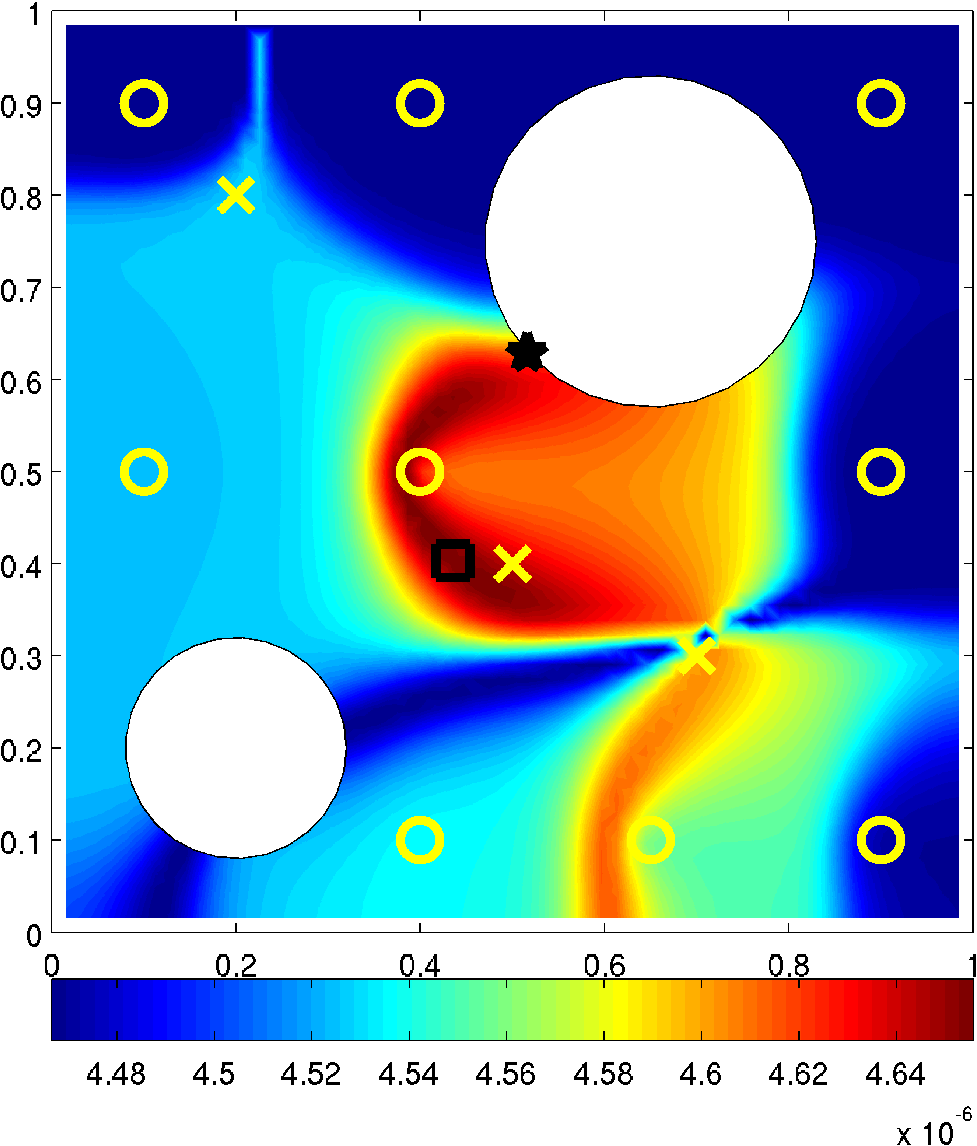}
\includegraphics[width=0.31\textwidth]{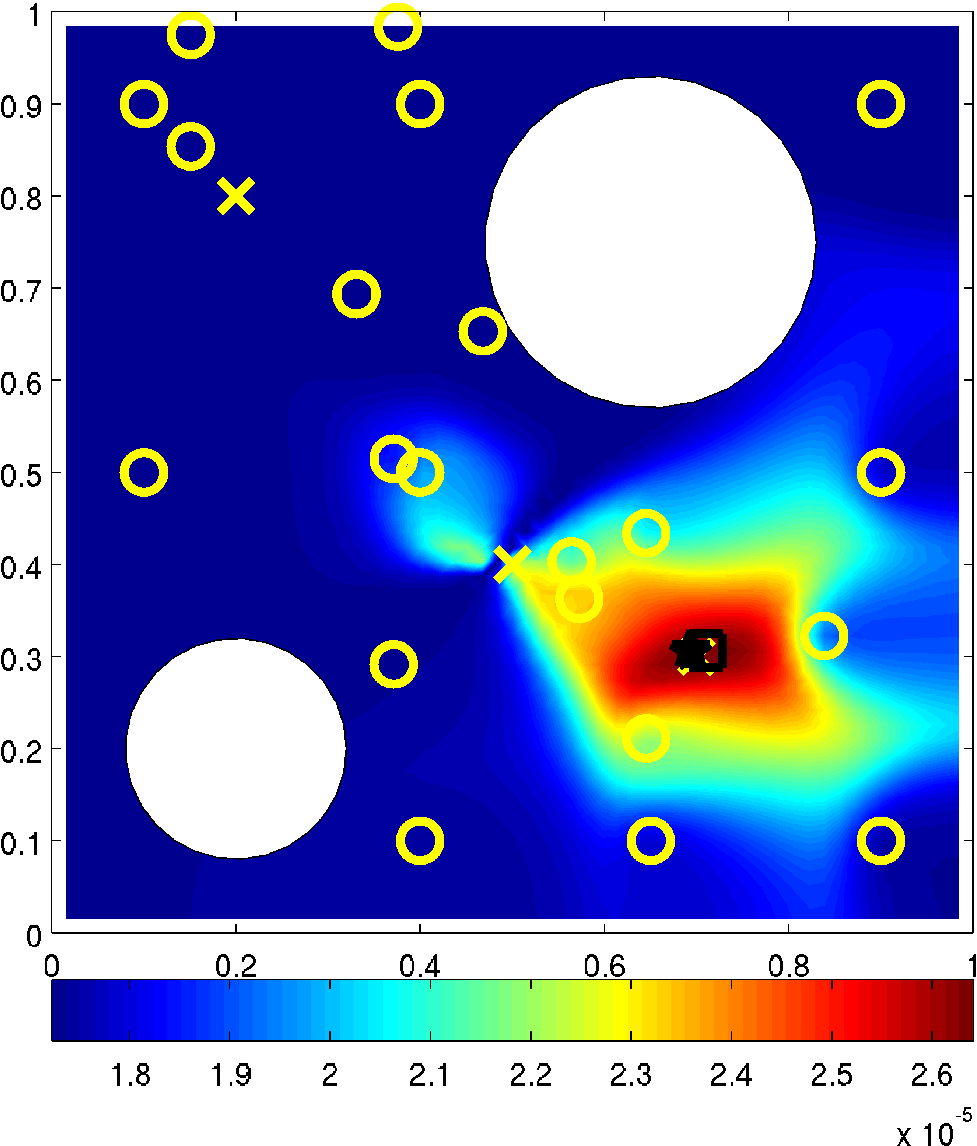} \hskip0.02\textwidth
\includegraphics[width=0.31\textwidth]{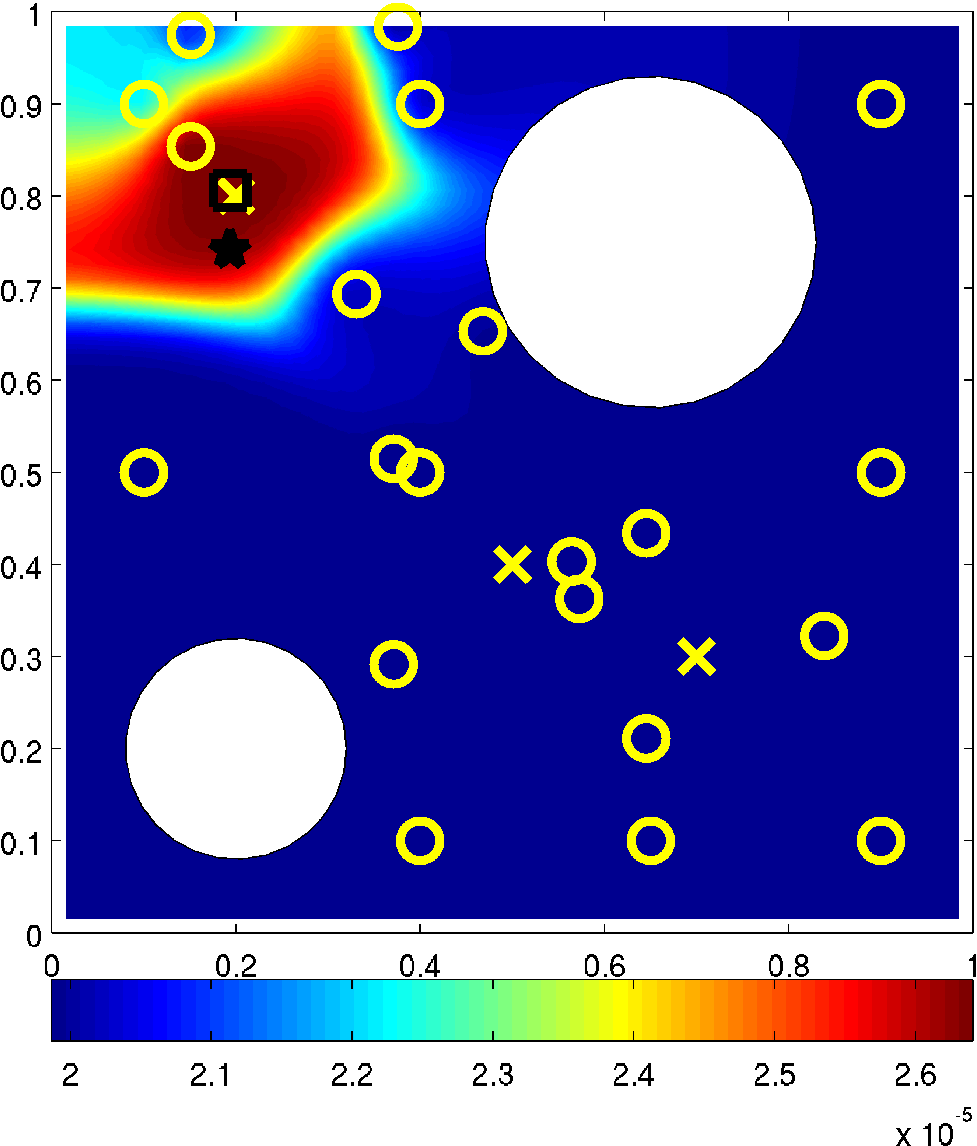} \hskip0.02\textwidth
\includegraphics[width=0.31\textwidth]{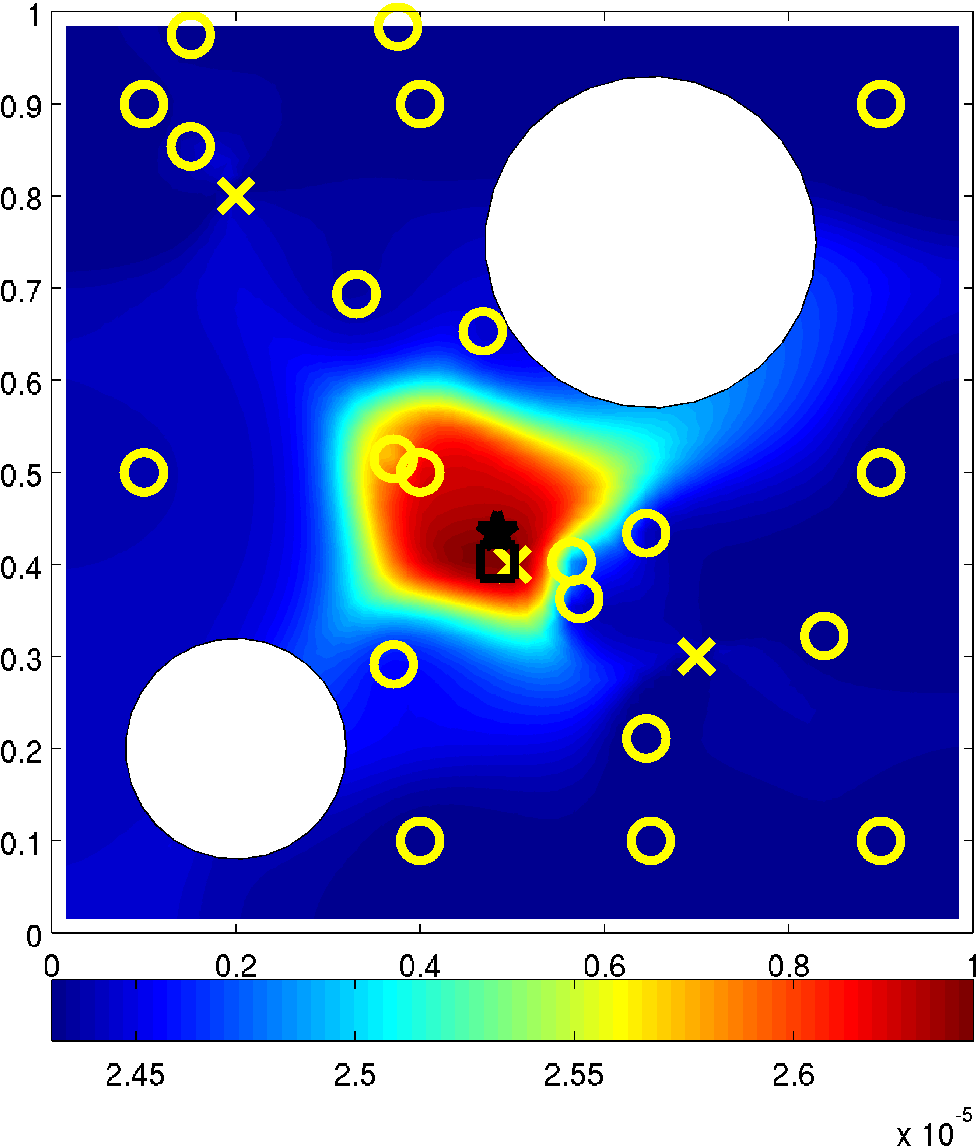}
\caption{Imaging functionals $J(\boldsymbol r, \boldsymbol y^2, \boldsymbol y^3)$ (left column),
$J(\boldsymbol y^1, \boldsymbol r, \boldsymbol y^3)$ (middle column) and
$J(\boldsymbol y^1, \boldsymbol y^2, \boldsymbol r)$ (right column) from (\ref{eqn:maxjj}) for all 
$\boldsymbol r \in \Omega$ evaluated at the final value of $(\boldsymbol y^1, \boldsymbol y^2, \boldsymbol y^3)$ 
given by Algorithm \ref{alg:fwdadj}.
Top row: initial run with $N_m=9$ measurements. Bottom row: subsequent run with measurements added adaptively
by Algorithm \ref{alg:adapmeas} ($N_m = 21$). True source locations are yellow $\times$, measurement locations 
are yellow $\circ$, estimated source position $(\boldsymbol y^1, \boldsymbol y^2, \boldsymbol y^3)$ - 
maximum of the imaging functional is black $\square$, initial guess from Algorithm \ref{alg:initguess} is black $\star$.
}
\label{fig:srcd3}
\end{center}
\end{figure}

In this section we study the identification of a known number of time independent sources from integrated 
measurements in a three component system from Section \ref{sec:threesystem}. We consider three cases $N_s = 1,2,3$ in 
Figures \ref{fig:srcd1}, \ref{fig:srcd2} and \ref{fig:srcd3} respectively. To demonstrate the adaptive measurement 
placement we begin by choosing the smallest number of measurements $N_m = 3 N_s$ that yields a formally determined 
system (\ref{eqn:adjvector}). Then we add more measurements using Algorithm \ref{alg:levelsetmeas} in the case
$N_s = 1$ and Algorithm \ref{alg:adapmeas} for $N_s = 2,3$. Then we run Algorithm \ref{alg:fwdadj} again with
the adaptively added measurements. For the purposes of visualization for the $j^{th}$ source we fix $k^{th}$ source 
locations given by Algorithm \ref{alg:fwdadj} for all $k \neq j$ and evaluate the functional (\ref{eqn:maxjj}) 
for all possible locations of the $j^{th}$ source in $\Omega$. This is similar to step (ii) of Algorithm 
\ref{alg:derivfree}. Obviously, the maximum of the functional corresponds to the source location estimated by 
Algorithm \ref{alg:fwdadj}. Doing so allows us to visualize the sensitivity of the objective of (\ref{eqn:optimize})
with respect to the location of the $j^{th}$ source and also how it changes when more measurements are added 
adaptively. 

\begin{table}
\caption{\label{tab:srcd3}True and reconstructed source intensities $a_j$ and locations $\boldsymbol y^j$,
and relative location errors $E$ for the case of three sources ($5\%$ noise in the data).}
\begin{tabular}{@{}cccccccc}
\br
Case & $a_1$ & $a_2$ & $a_3$ & $\boldsymbol y^1$ & $\boldsymbol y^2$ & $\boldsymbol y^3$ & $E$ \\
\mr
True       & 10.00 & 7.00 & 5.00 & (0.70, 0.30) & (0.20, 0.80) & (0.50, 0.40) & -- \\
$N_m = 9$  & 13.06 & 21.42 & 4.22 & (0.70, 0.32) & (0.22, 0.98) & (0.43, 0.40) & 0.09 \\
$N_m = 21$ & 9.35 & 7.26 & 5.31 & (0.70, 0.30) & (0.19, 0.80) & (0.48, 0.40) & 0.01 \\
\br
\end{tabular}
\end{table}

Detection of a single source is rather robust, so even in the presence of measurement noise the source is identified
almost exactly with just three measurements, as is illustrated in the left plot in Figure \ref{fig:srcd1}. However, the 
imaging functional has a large plateau around the true source location due to lack of downwind measurements. 
Algorithm \ref{alg:levelsetmeas} performs as expected by adding a new measurement downwind and the imaging functional
becomes much better localized as shown in the right plot in Figure \ref{fig:srcd1}.

In Figure \ref{fig:srcd2} we observe that the objective is not convex (the functional $J$ is not concave). 
Also in Figure \ref{fig:srcd2} for $N_m = 9$ measurements it is clear that the objective can develop narrow 
valleys (ridges of $J$) and become multimodal. This makes the multiple source identification problem difficult to solve, 
and we observe that in the presence of noise the estimated source location may differ from the true one if too few 
measurements are used. In particular, we see in top rows of Figures \ref{fig:srcd2} and \ref{fig:srcd3} that the 
estimated locations are off for the sources for which the objective has a large plateau around the true location.


In Figures \ref{fig:srcd2} and \ref{fig:srcd3} we only show the estimated locations of the sources. The corresponding
source intensities (and the numerical values for the locations) are given in Tables \ref{tab:srcd2} and \ref{tab:srcd3}
for the cases $N_s = 2$ and $N_s = 3$ respectively. From the presented data we observe that the least squares 
estimate (\ref{eqn:leastsqa}) of source intensities is quite sensitive to the estimate of the source locations.
When few measurements are used and the estimates of the locations are not accurate enough the estimated intensities 
differ significantly from the true values. However, when more measurements are added adaptively, the intensity
estimates improve greatly. Note that this limitation of our method comes from the fact that in (\ref{eqn:leastsqa}) 
we eliminate the source intensities from the optimization variables. If we have some a priory knowledge about the
intensities (i.e. the bounds) we can retain $\boldsymbol a$ as an optimization variable in (\ref{eqn:optimize})
and enforce our a priori knowledge as a constraint. This comes at a price of enlarging the space of optimization
variables, which makes the optimization problem harder to solve.

\subsection{Source identification for an unknown number of sources}
\label{sec:numunknown}

\begin{figure}[t!]
\begin{center}
\includegraphics[width=0.30\textwidth]{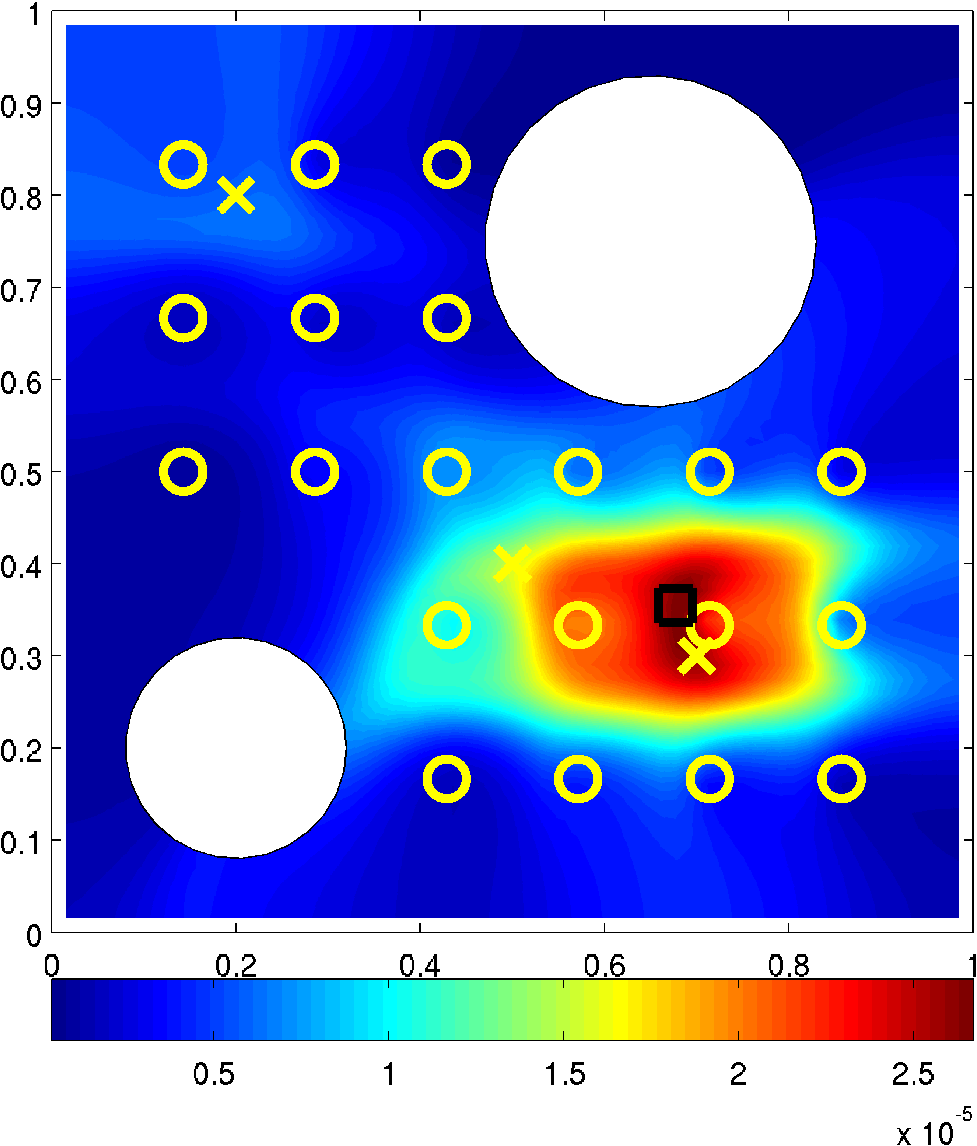} \hskip0.02\textwidth
\includegraphics[width=0.30\textwidth]{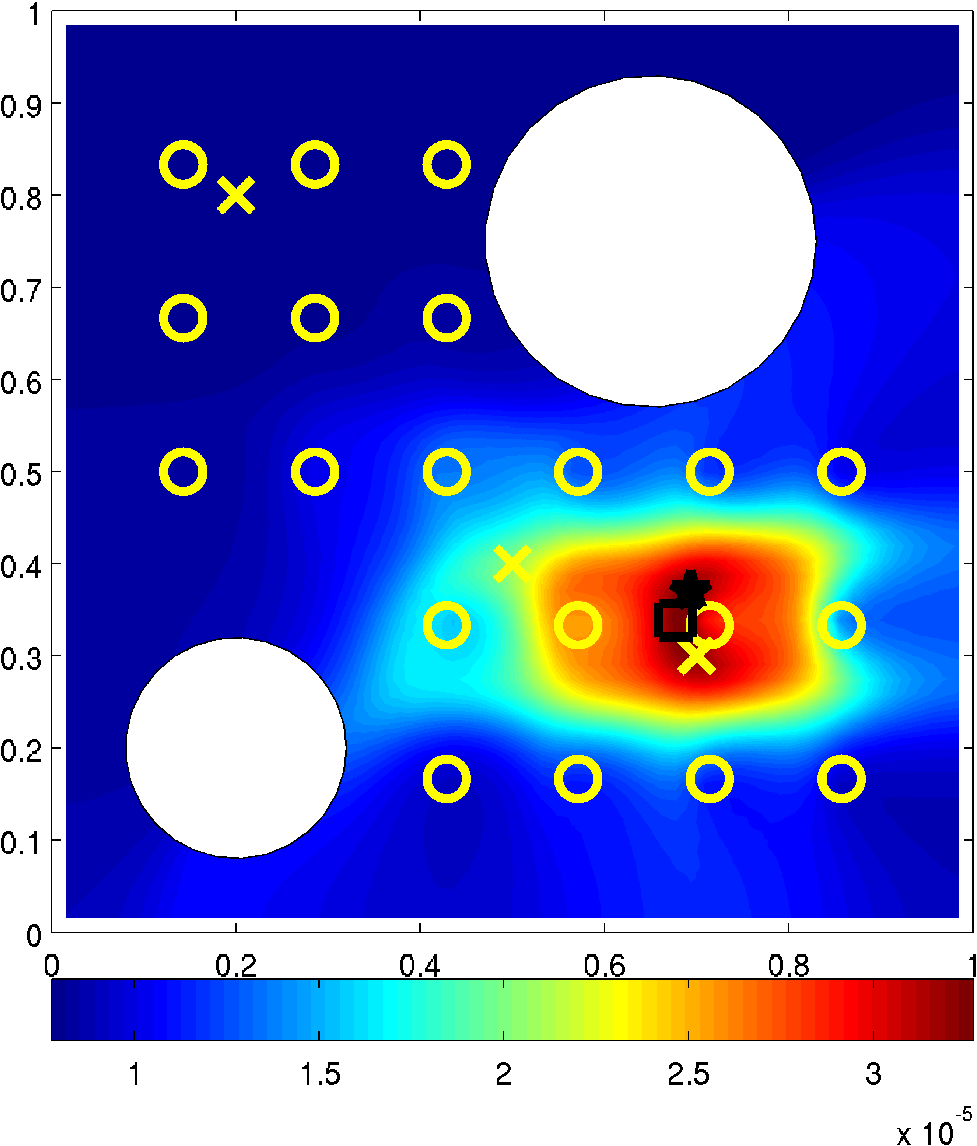} \hskip0.02\textwidth
\includegraphics[width=0.30\textwidth]{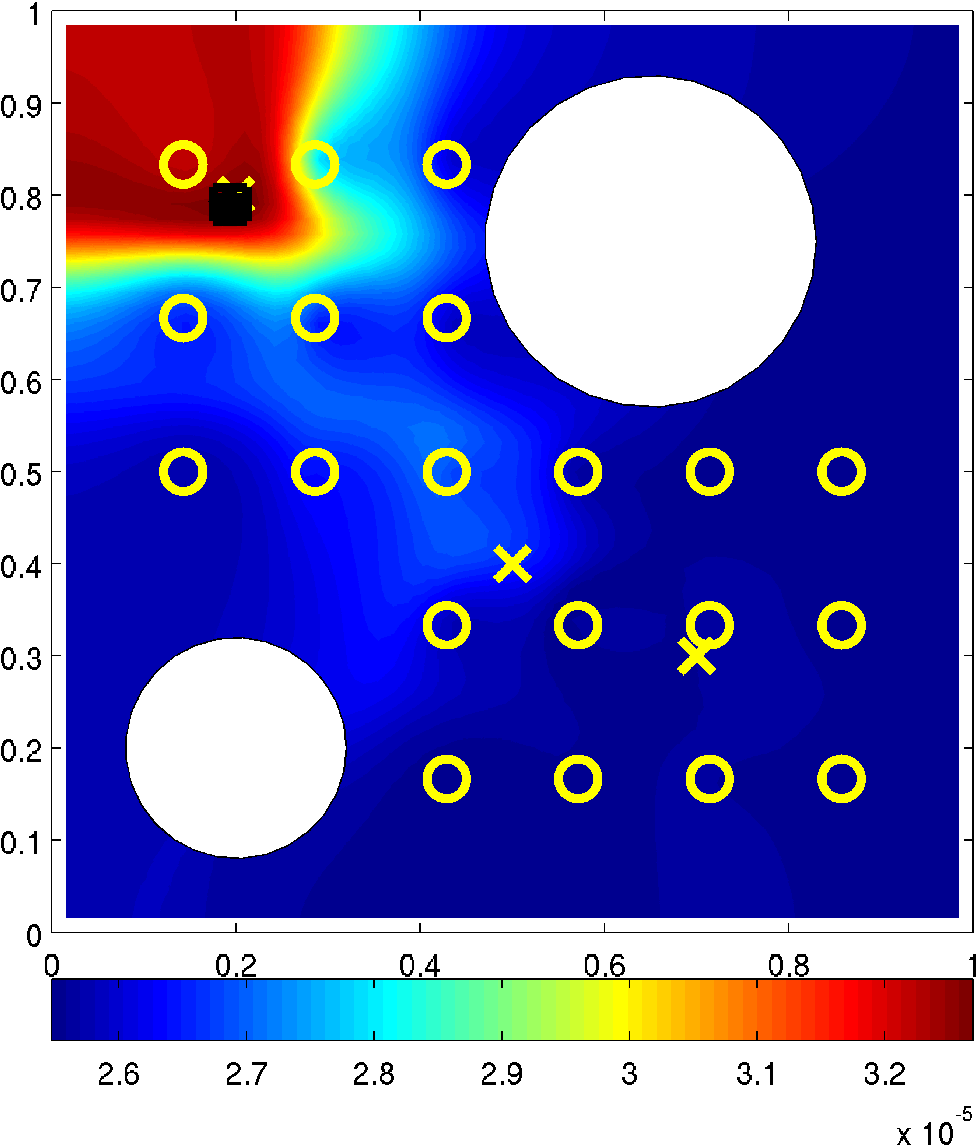}
\includegraphics[width=0.30\textwidth]{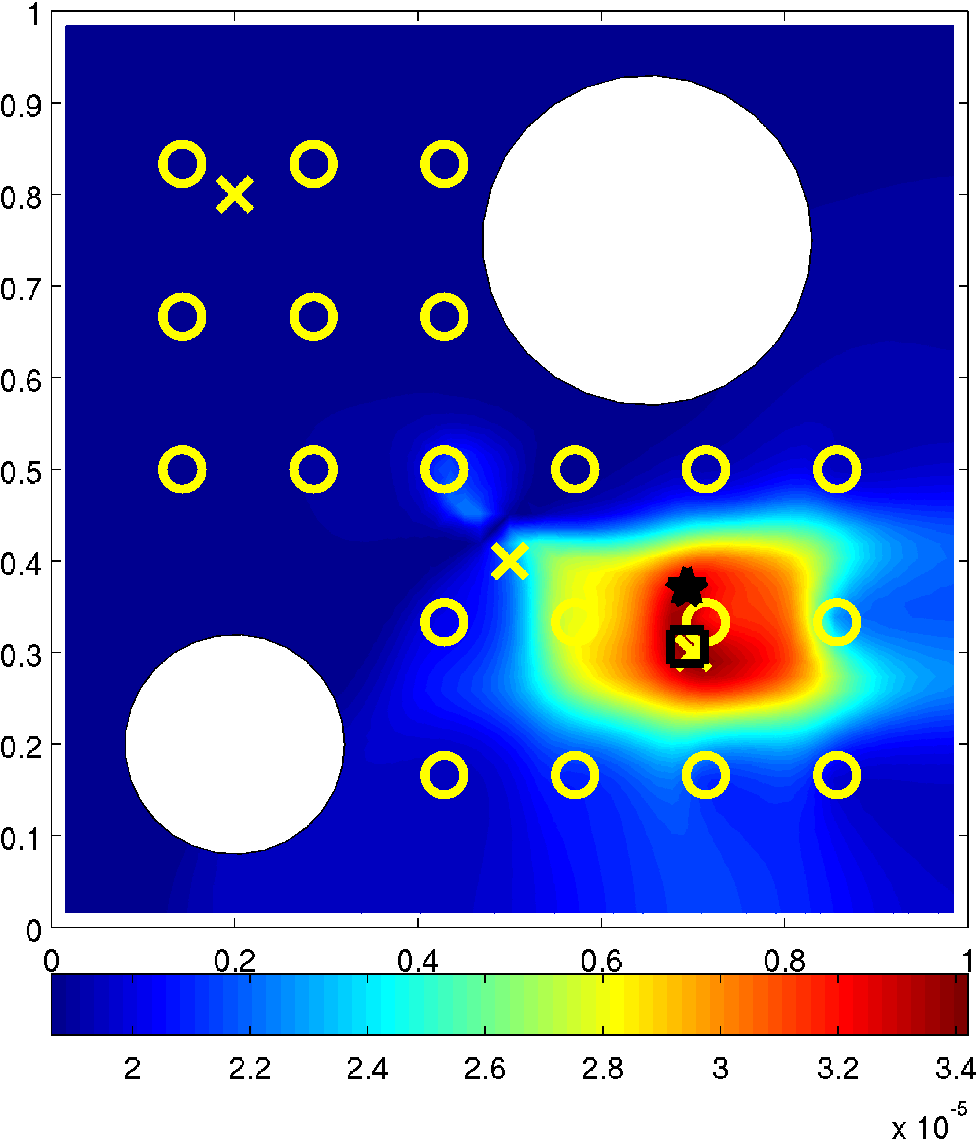} \hskip0.02\textwidth
\includegraphics[width=0.30\textwidth]{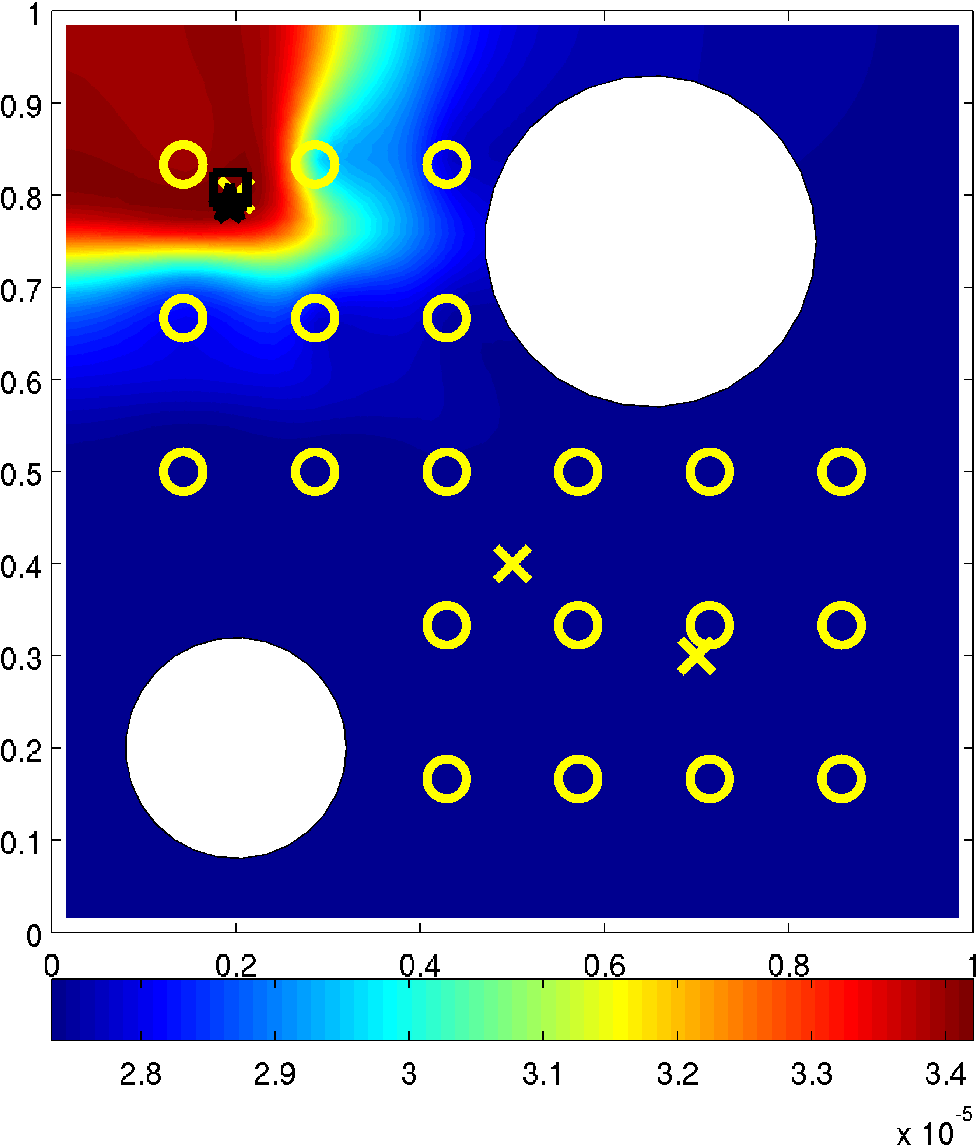} \hskip0.02\textwidth
\includegraphics[width=0.30\textwidth]{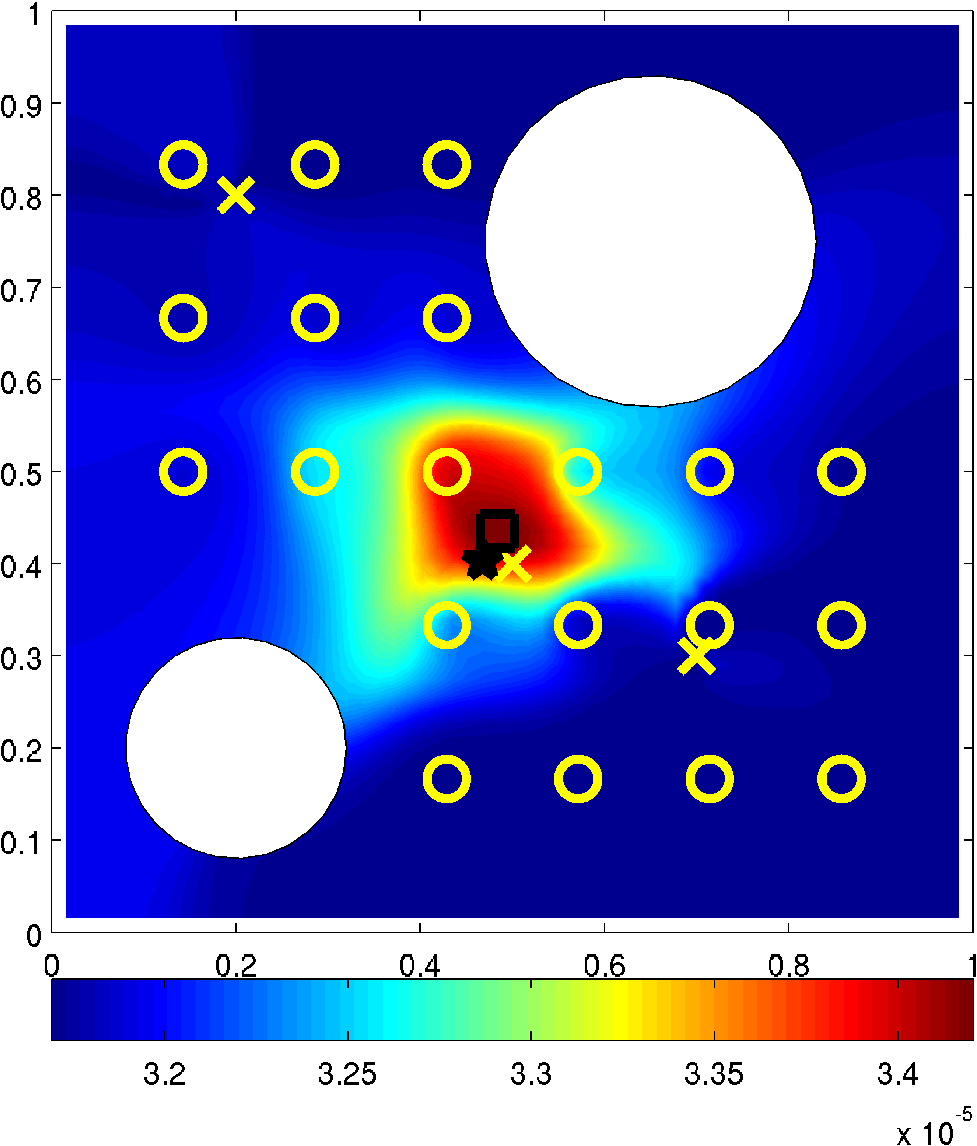}
\includegraphics[width=0.23\textwidth]{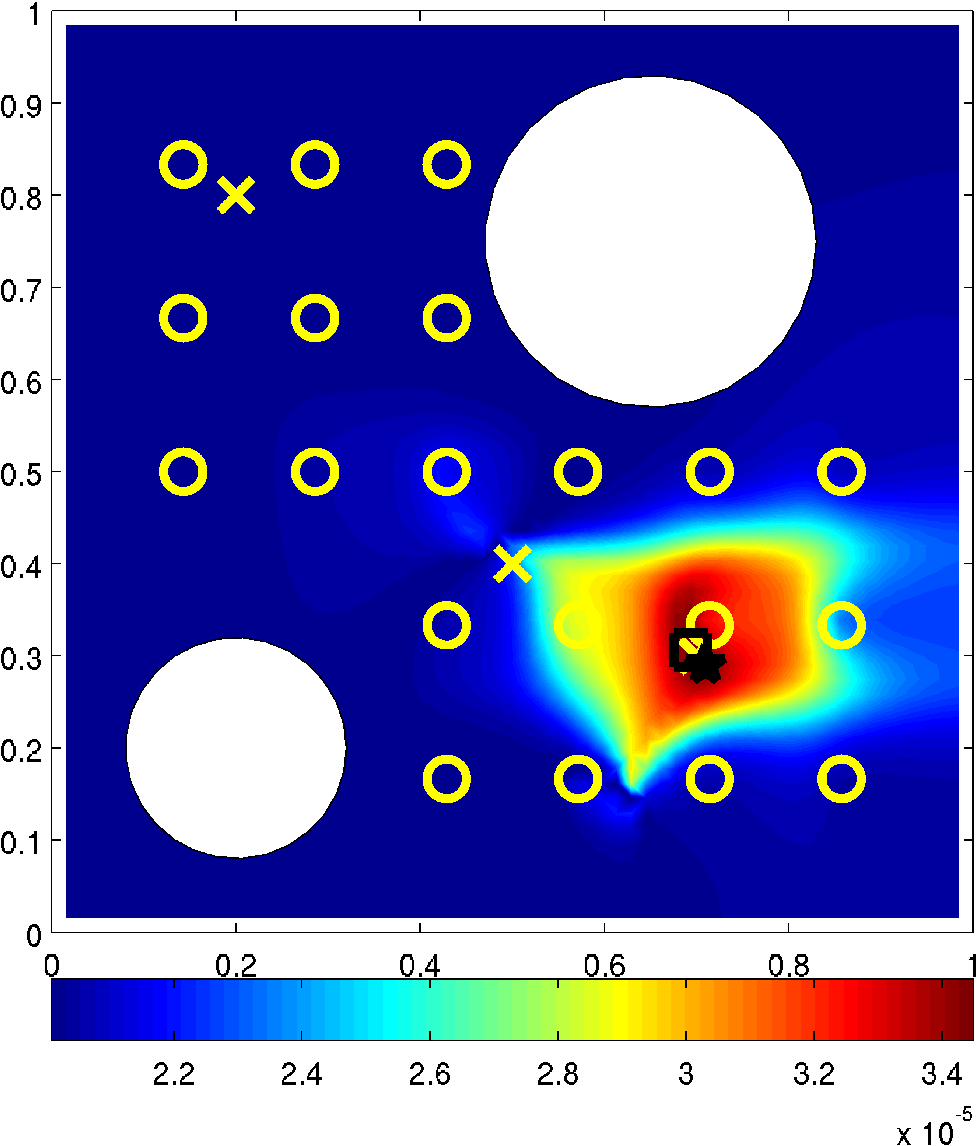} \hskip0.01\textwidth
\includegraphics[width=0.23\textwidth]{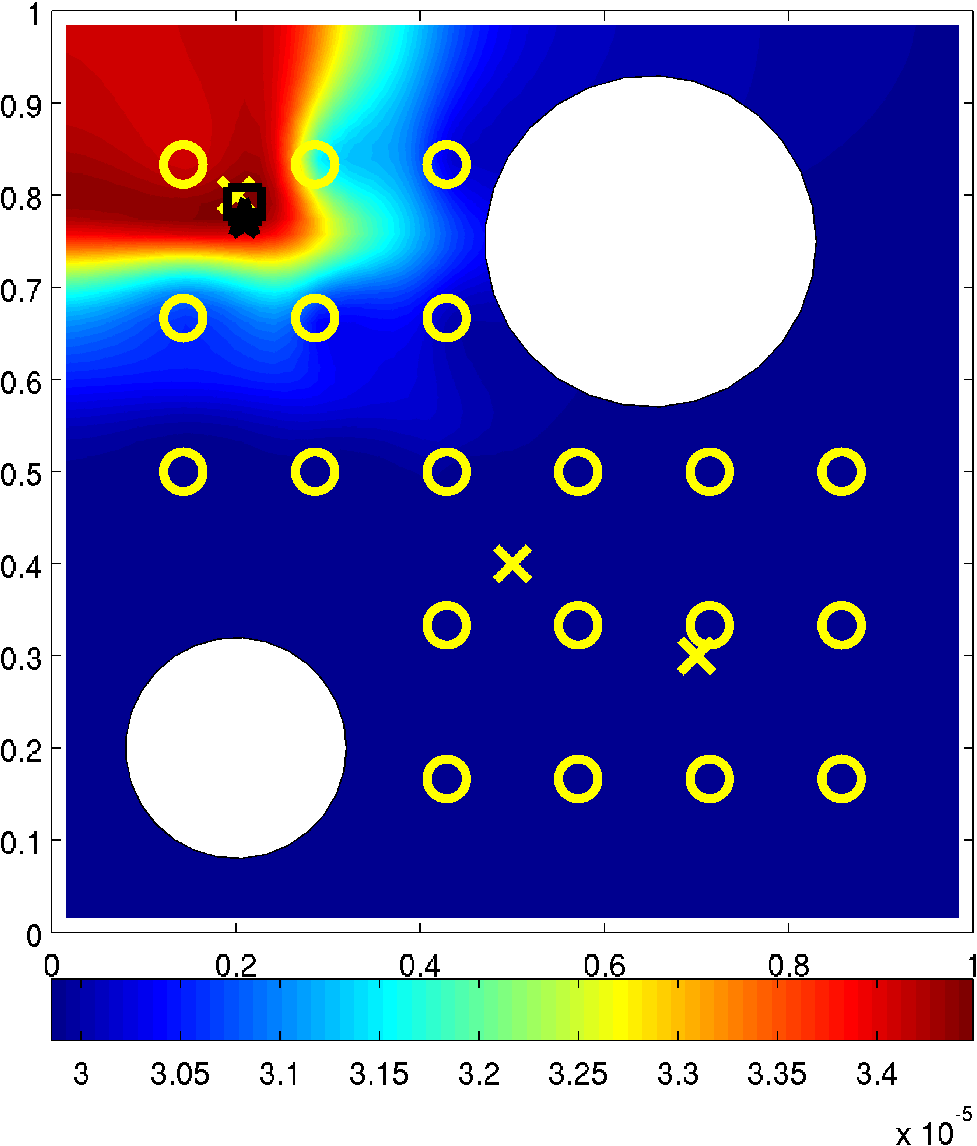} \hskip0.01\textwidth
\includegraphics[width=0.23\textwidth]{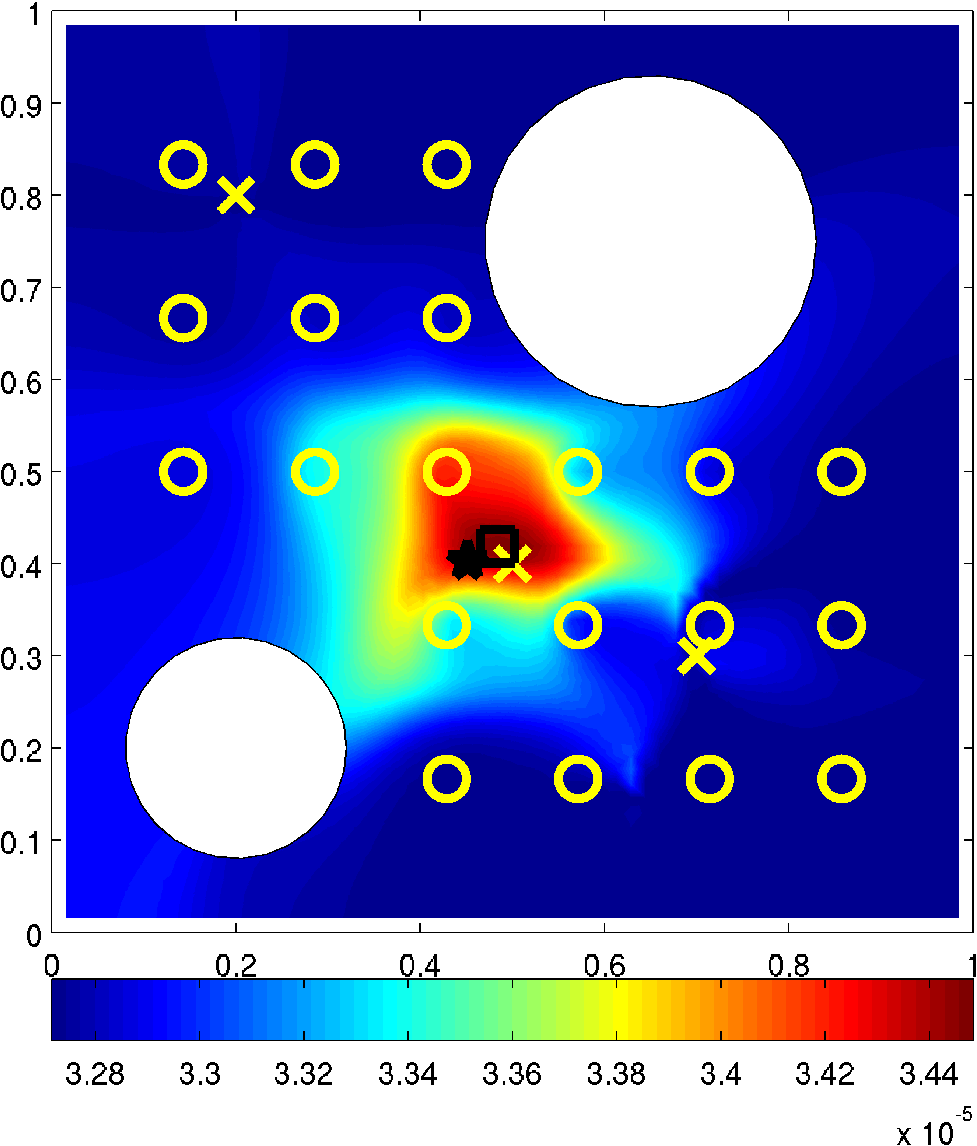} \hskip0.01\textwidth
\includegraphics[width=0.23\textwidth]{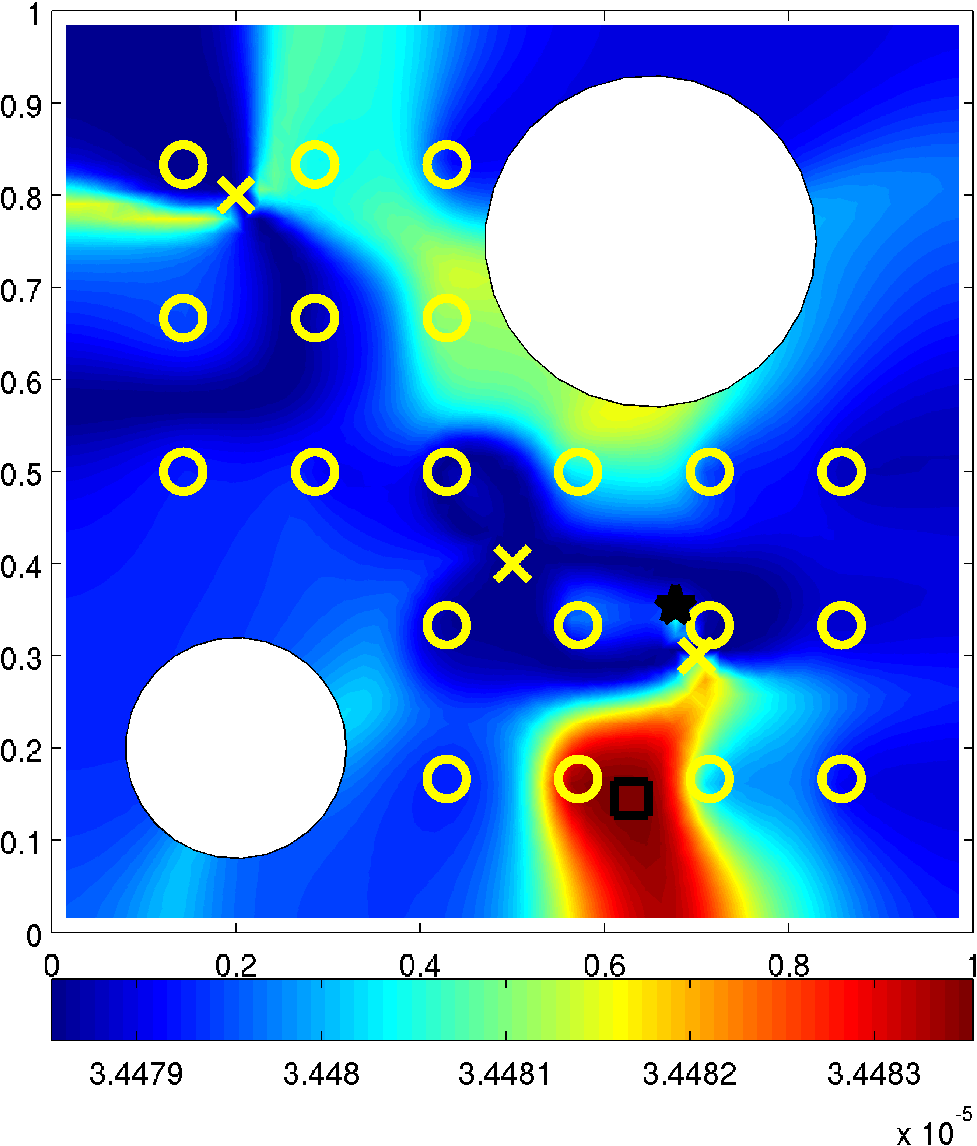} 
\caption{Unknown number of sources, the true value is $N_s=3$. Imaging functionals evaluated at the final value of 
$(\boldsymbol y^1, \ldots, \boldsymbol y^{N_s^*})$ given by Algorithm \ref{alg:fwdadj}.
Top row: $N_s^*=1$ (leftmost) and $N_s^*=2$ (middle and right). Middle row: $N_s^*=3$. Bottom row: $N_s^*=4$. 
Number of measurements $N_m = 20$ for all $N_s^*$. True source locations are yellow $\times$, measurement locations 
are yellow $\circ$, estimated source positions $(\boldsymbol y^1, \ldots, \boldsymbol y^{N_s^*})$ - maxima of the 
corresponding imaging functional are black $\square$, initial guesses from Algorithm \ref{alg:initguess} are black 
$\star$.}
\label{fig:srcdnum}
\end{center}
\end{figure}

Let us now consider source identification in the case when a true value of $N_s$ is unknown. Similarly to the previous 
section we identify time independent sources from integrated measurements. To identify the sources
and their number we use the procedure from Section \ref{sec:measposnumdet}. To simplify the exposition in this section 
we do not add the measurements adaptively. Instead we use a predetermined large number of measurements for all trial 
values of $N_s^*$. The measurements are distributed somewhat uniformly in $\Omega$, as shown in Figure \ref{fig:srcdnum}.

\begin{table}
\caption{\label{tab:srcdnum}True and reconstructed source intensities $a_j$ and locations $\boldsymbol y^j$ 
for the case of unknown number of sources ($5\%$ noise in the data).}
\begin{tabular}{@{}ccccccccc}
\br
Case & $a_1$ & $a_2$ & $a_3$ & $a_4$ & $\boldsymbol y^1$ & $\boldsymbol y^2$ & $\boldsymbol y^3$ & $\boldsymbol y^4$ \\
\mr
\parbox{0.12\textwidth}{~~~True\\ ($N_s=3$)} & 10.00 & 7.00 & 5.00 & --    & (0.70, 0.30) & (0.20, 0.80) & (0.50, 0.40) & -- \\
$N_s^* = 1$  & 12.29 &  --  &  --  & --    & (0.67, 0.35) &    --        &   --         & -- \\
$N_s^* = 2$  & 12.12 & 8.13 &  --  & --    & (0.67, 0.33) & (0.19, 0.79) &   --         & -- \\
$N_s^* = 3$  & 10.44 & 6.83 & 4.44 & --    & (0.69, 0.30) & (0.19, 0.80) & (0.48, 0.43) & -- \\
$N_s^* = 4$  & 11.30 & 6.42 & 4.12 & -0.17 & (0.69, 0.30) & (0.20, 0.79) & (0.48, 0.41) & (0.62, 0.14) \\
\br
\end{tabular}
\end{table}

We set $N_s = 3$ and we perform four trials $N_s^* = 1,2,3,4$. The results of these trials along with the true source 
parameters are given in Table \ref{tab:srcdnum} and are visualized in Figure \ref{fig:srcdnum}. 
We observe that as we increase the the trial number 
$N_s^*$ Algorithm \ref{alg:fwdadj} starts to ``notice'' the sources with smaller intensity. At the first step
$N_s^* = 1$ it picks the dominant source with $a_1 = 10$. At the second step it notices the presence of the 
source with $a_2 = 7$. Note that the locations and intensities of the first two sources are not determined
exactly, because the objective of (\ref{eqn:maximize}) is different from the true one unless $N_s^* = N_s$. 
However, the estimates of the locations and intensities of the first two sources while not exact are quite
accurate, as can be observed in the first row of Figure \ref{fig:srcdnum} and also in Table \ref{tab:srcdnum}.


Finally, when we reach $N_s^* = N_s = 3$ the method identifies all three sources quite reliably given the level of
noise present. As we go one step further $N_s^* = 4$ Algorithm \ref{alg:fwdadj} recovers a spurious source
with a negative intensity, which we use as a stopping criterion. We conclude that the true sources were recovered 
in the previous step and the true number of sources is $N_s = 3$.
Note that while a spurious source appears in the case $N_s^* = 4$, the method gives a good estimate of the 
locations and intensities of the three sources. We observed such behavior for many realizations of the noise, so
the results presented in Figure \ref{fig:srcdnum} and Table \ref{tab:srcdnum} are representative of the general
performance of the method.

\subsection{Time dependent source identification}
\label{sec:timesrcid}

In the numerical examples considered above we used time independent sources that are active for all $t$ in $[0,T]$. 
In this section we apply Algorithm \ref{alg:fwdadj} to identify a single time dependent source, which is a point 
source in both space and time. For simplicity of visualization we first consider in Section \ref{sec:time1d} a problem
in one spatial dimension. This allows us to plot the imaging functional for all space and time locations. In section
\ref{sec:time2d} we consider the example in two dimensions for the three component chemical system.

\subsubsection{One dimensional case}
\label{sec:time1d}

\begin{figure}[t!]
\begin{center}
\includegraphics[width=0.48\textwidth]{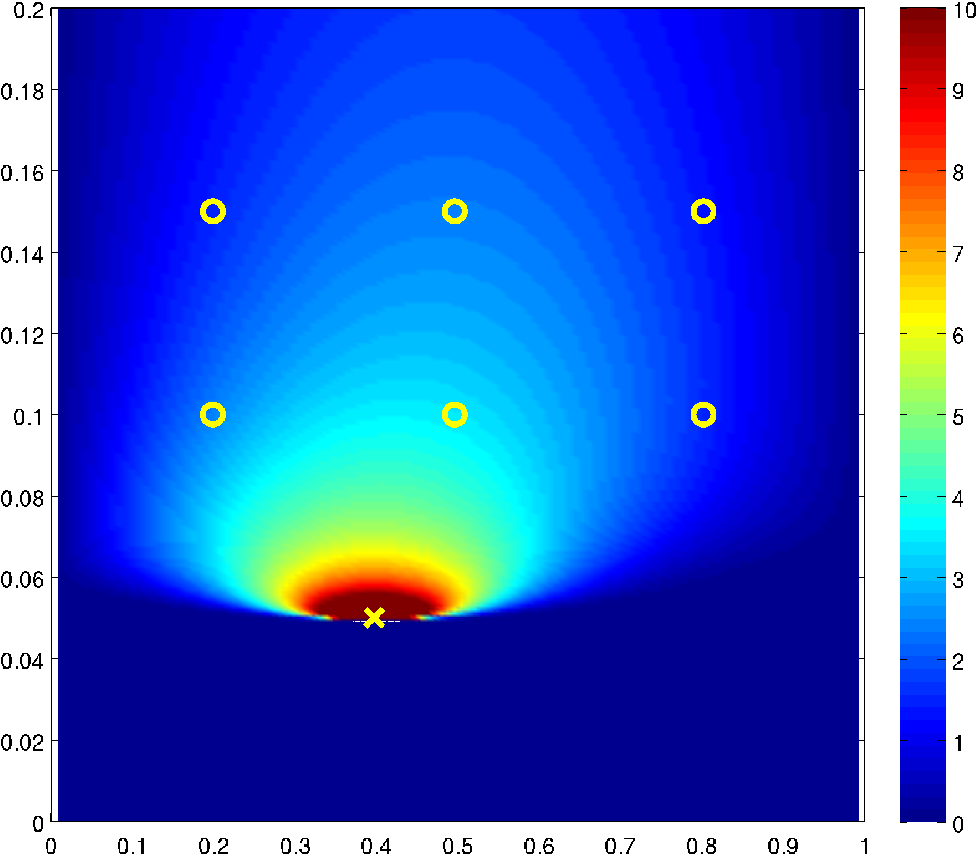} \hskip0.03\textwidth
\includegraphics[width=0.48\textwidth]{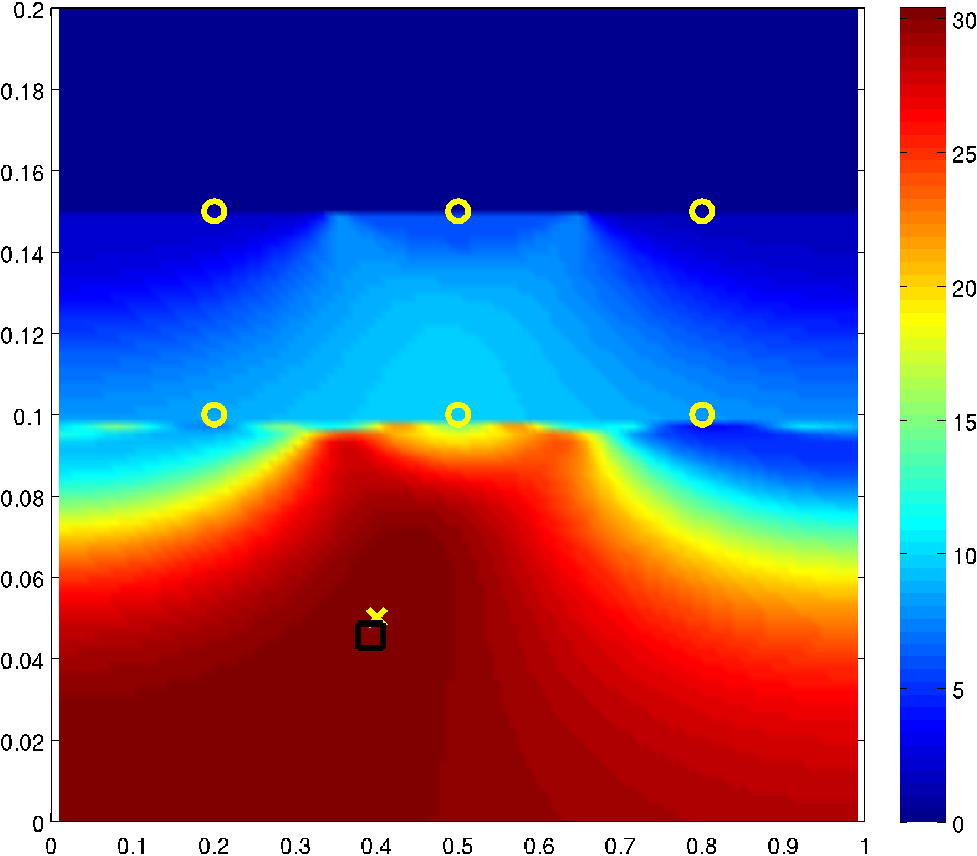}
\caption{Identification of a single time dependent source from instantaneous measurements in one dimension with 
$1\%$ noise in the data. 
Left: forward problem solution $u(x,t)$.  
Right: imaging functional $\boldsymbol J(\boldsymbol s)$ for $\boldsymbol s = (x,t) \in [0,1] \times [0,0.2]$. 
Horizontal axis is $x$, vertical axis is $t$.
True source location is yellow $\times$, measurement locations are yellow $\circ$, 
source position estimated by Algorithm \ref{alg:fwdadj} is black $\square$. True source
parameters $(a, y, \tau)$ are $(3, 0.4, 0.05)$, estimated are $(3.178, 0.392, 0.045)$.}
\label{fig:timedep}
\end{center}
\end{figure}

Let us consider a scalar forward problem of the form (\ref{eqn:usystem}) with $\epsilon = 1$, 
$\boldsymbol w = 0$, $\boldsymbol L = 5$, $\boldsymbol Q(\boldsymbol u) = - u$, $\Omega = [0,1]$ and $T = 0.2$.
A single source of the form
\begin{equation}
f (x, t) = a \delta(t-\tau) \delta(x-y)
\label{eqn:timesrc}
\end{equation}
is to be determined, where $a=3$, $\tau = 0.05$ and $y = 0.4$.

Similarly to the two dimensional case we use a finite difference scheme in space and an exponential integrator in time.
To avoid committing an inverse crime we use a fine grid to compute the forward solution for simulating the data
with $200$ grid steps in $x$ and $120$ time steps. A coarser grid is used in Algorithm \ref{alg:fwdadj} to
identify the source with $101$ steps in both spatial and temporal variables.

We observed from our numerical experiments that identification of time dependent sources is more sensitive to noise
and numerical errors than the identification of sources in examples in sections \ref{sec:idmultiadap} and 
\ref{sec:numunknown}. Thus, for stable source identification we need to use more measurements than is required to 
just make the system (\ref{eqn:adjvector}) formally determined. The source in (\ref{eqn:timesrc}) is determined by 
three parameters, but we use six measurements for our numerical example. We make measurements at spatial locations 
$0.2$, $0.5$ and $0.8$ at two time instants $0.1$ and $0.15$ for a total of $6$ measurements. 

The results of the forward simulation and source identification by Algorithm \ref{alg:fwdadj} are shown in 
Figure \ref{fig:timedep}. We observe that both the source location and its intensity were identified reasonably well
given the noisy data. The plot of the imaging functional $\boldsymbol J (\boldsymbol s)$ explains why the time 
dependent source identification problem is more difficult than the previously considered examples. The imaging functional
has a large plateau surrounding the true source location, which decreases the discriminatory power of the method. For
higher noise levels we observed that the estimated source location ends up somewhere on this plateau far away from the
true source position.

\subsubsection{Two dimensional case}
\label{sec:time2d}

\begin{figure}[t!]
\begin{center}
\includegraphics[width=0.32\textwidth]{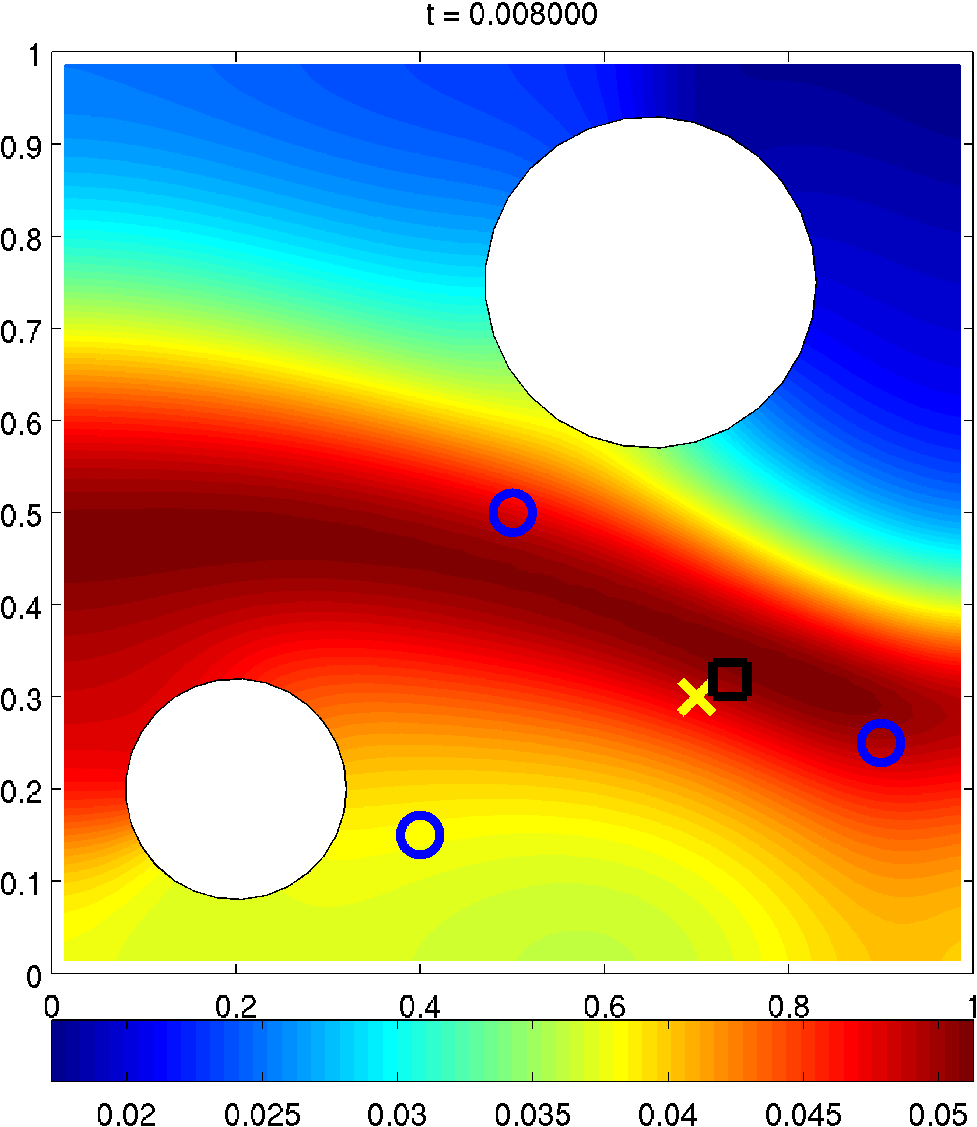} \hskip0.01\textwidth
\includegraphics[width=0.32\textwidth]{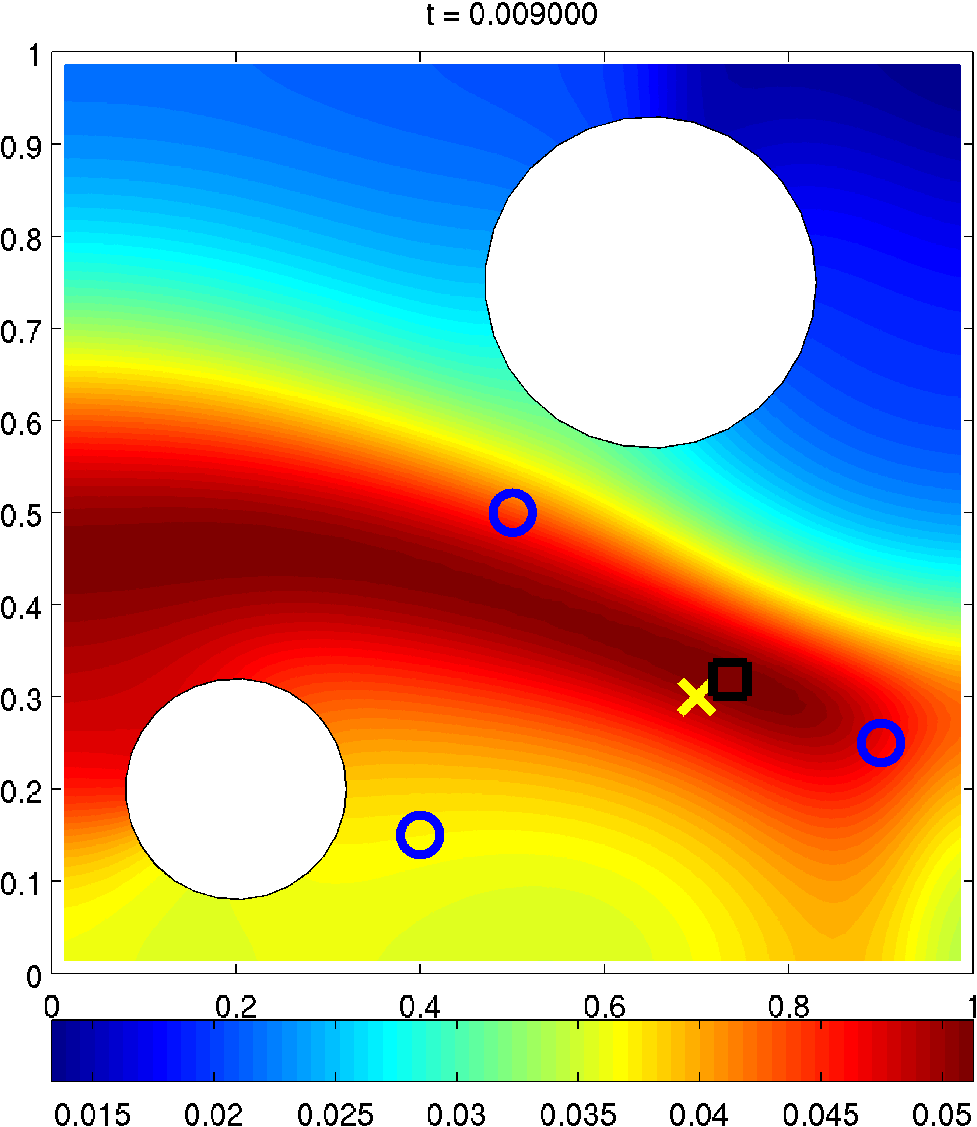} \hskip0.01\textwidth
\includegraphics[width=0.32\textwidth]{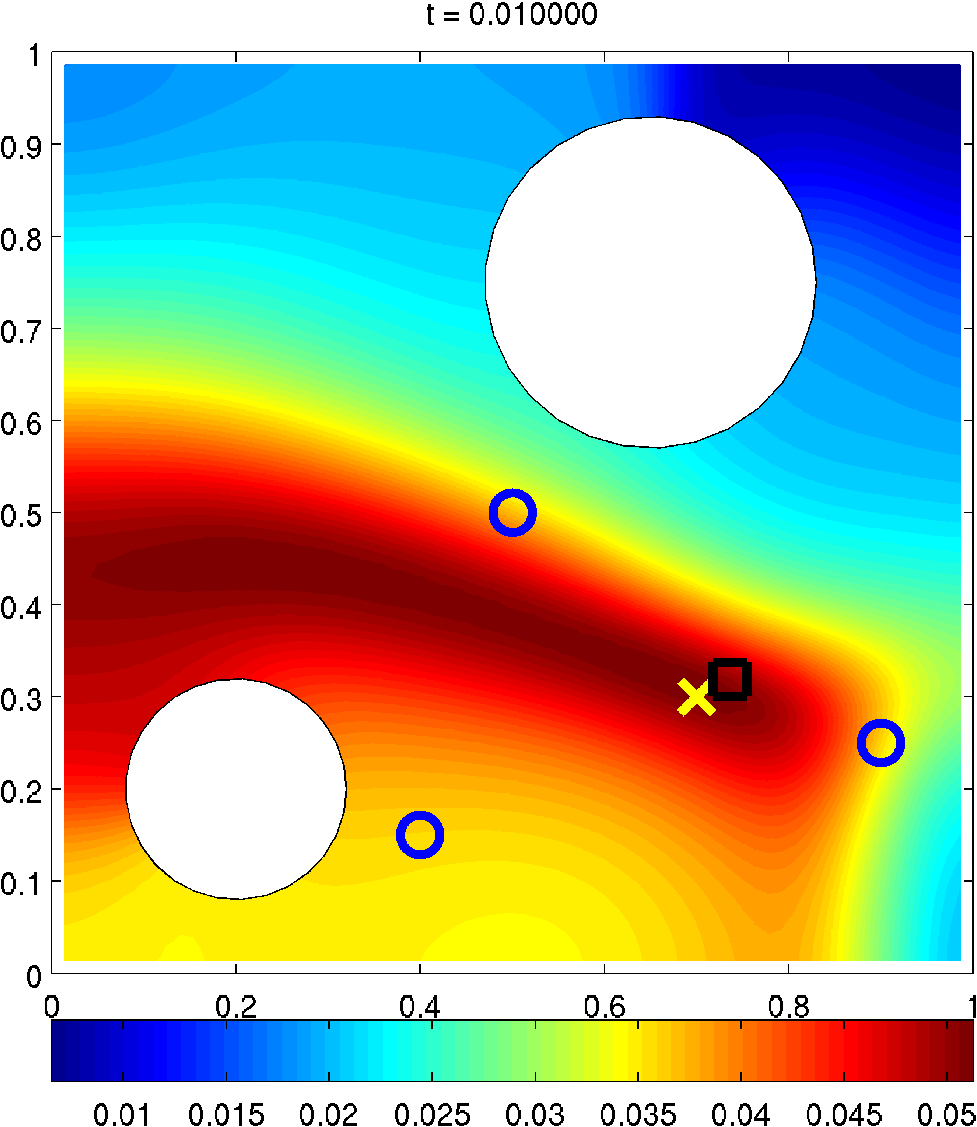}
\caption{Identification of a single time dependent source from instantaneous measurements in two dimensions with 
$1\%$ noise in the data. 
Left to right: slices of the imaging functional $J(\boldsymbol s)$, $\boldsymbol s = (\boldsymbol y, t)$ for the three values
of $t = 0.008, 0.009, 0.010$ and $\boldsymbol y \in \Omega$. True source spatial location is yellow $\times$, 
measurement spatial locations are blue $\circ$. Source spatial position estimated by Algorithm \ref{alg:fwdadj} is black $\square$.
True source parameters $(a,x,y,\tau)$ are $(10, 0.7, 0.3, 0.010)$, estimated are $(11.18, 0.73, 0.31, 0.009)$.}
\label{fig:timedep2d}
\end{center}
\end{figure}

Here we present the results of time dependent source identification for the three component chemical system in two dimensions
from instantaneous measurements. As we observed in Section \ref{sec:time1d} such source detection is more difficult than
the cases considered in sections \ref{sec:idmultiadap} and \ref{sec:numunknown}, so for stable identification we 
reduced the non-linearity of the system by taking smaller reaction rates $k_1 = 100$ and $k_2 = 200$. Higher
reaction rates leading to stiffer system can be handled using more efficient numerical schemes, for example 
\cite{engquist2005heterogeneous}. However, proper numerical treatment of stiff systems is out of the scope of this work,
so for convenience we work with reduced reaction rates in this section.

We simulate the system up to $T=0.03$, which is the time when the system is still in transient behavior. The source goes 
off at $\tau = 0.01$ and we make two sets of three point measurements each at instants $\theta = 0.015$ and 
$\theta = 0.020$ for a total of six measurements.

In Figure \ref{fig:timedep2d} we show three slices of the imaging functional $J(\boldsymbol s)$ at time instants adjacent 
to the temporal source location estimated by Algorithm \ref{alg:fwdadj}. We observe that the imaging functional has a 
narrow ridge with a plateau on top, which can make source identification difficult similarly to the one dimensional
case, where in the presence of noise Algorithm \ref{alg:fwdadj} can get stuck far away from the true source location.

\section{Conclusions and future work}
\label{sec:conclude}

We presented here a method for source identification in non-linear time dependent advection-diffusion-reaction systems.
We also provided the results of extensive numerical experiments that suggest that our method performs well in the 
presence of noise in the data and/or uncertainty in the number of sources present. The numerical experiments also
show that the method's performance can be further improved by adaptively adding more measurements using the proposed
strategy. 

The following topics of future study can be proposed. First, determining the conditions under which Algorithm
\ref{alg:fwdadj} converges and proving the convergence. The analysis is complicated by the lack of regularity
of solutions in the presence of point sources and by the coupling between the forward iteration and source estimation
at each step of the algorithm. 

Second, the study of the case where only a partial knowledge of domain $\Omega$ is assumed. In this case both the 
sources and the obstacles need to be determined. A method proposed in \cite{burger2009discovering} solves the linear 
case by using the comparison results for elliptic equations. Since similar results hold for non-linear parabolic 
systems, it should be possible to extend the method in \cite{burger2009discovering} to the setting considered here. 

Third, in this paper we assumed that all system parameters such as reaction rates, advection field and diffusion 
coefficients are known. In reality these parameters are estimated from some other measurements and thus are prone to 
inaccuracies. One may study the sensitivity of source identification with respect to uncertainties in the system 
parameters, or even try to estimate these parameters as a part of the source identification problem.

\section*{Acknowledgements}
\label{sec:acknowledge}

The research of Mamonov and Tsai was supported by NSF grants DMS-0914465 and DMS-0914840. The authors thank the anonymous
referees for valuable comments and suggestions that helped improve the manuscript.

\section*{References}
\bibliography{biblio} 
\bibliographystyle{plain}
\end{document}